\documentclass[sigconf, nonacm]{acmart}

\usepackage{algorithmic}
\usepackage{graphicx}
\usepackage{textcomp}
\usepackage{diagbox}
\usepackage{xcolor}
\usepackage[vlined,ruled]{algorithm2e}
\usepackage{bm}
\usepackage{multirow}
\usepackage{color,xcolor}
\usepackage{enumitem}
\usepackage{tikz}
\usepackage{subcaption}

\newcommand{\eps}{0.5}
\newif\ifshowproof

\showprooftrue

\usepackage{amsthm}
\newtheorem{definition}{Definition}

\newtheorem{property}{Property}
\newtheorem{theorem}{Theorem}

\newtheorem{lemma}{Lemma}

\newtheorem{corollary}{Corollary}
\newtheorem{example}{Example}

\newcommand{\PY}[1]{{\color{black}#1}}

\newcommand\vldbdoi{XX.XX/XXX.XX}
\newcommand\vldbpages{XXX-XXX}
\newcommand\vldbvolume{14}
\newcommand\vldbissue{1}
\newcommand\vldbyear{2025}
\newcommand\vldbauthors{\authors}
\newcommand\vldbtitle{\shorttitle} 
\newcommand\vldbavailabilityurl{https://github.com/Chen-Py/JoinREnum}
\newcommand\vldbpagestyle{plain} 

\begin{document}
\title{Towards Efficient Random-Order Enumeration for Join Queries}


\author{Pengyu Chen}
\email{pchen.research@gmail.com}
\affiliation{%
  \institution{Harbin Institute of Technology}
  \city{Harbin}
  \country{China}
}

\author{Zizheng Guo}
\email{zguo.research@gmail.com}
\affiliation{%
 \institution{Harbin Institute of Technology}
 \city{Harbin}
 \country{China}
}
\author{Jianwei Yang}
\email{yangjianwei006@cnpc.com.cn}
\affiliation{%
  \institution{Daqing Oilfield Digital \& Intelligent Technology Company}
  \city{Daqing}
  \country{China}
}
\author{Dongjing Miao}
\email{miaodongjing@hit.edu.cn}
\affiliation{%
 \institution{Harbin Institute of Technology}
 \city{Harbin}
 \country{China}
}







\begin{abstract}
In many data analysis pipelines, a basic and time-consuming process is to produce join results and feed them into downstream tasks. Numerous enumeration algorithms have been developed for this purpose. To be a statistically meaningful representation of the whole join result, the result tuples are required to be enumerated in uniformly random order. However, existing studies lack an efficient random-order enumeration algorithm with a worst-case runtime guarantee for (cyclic) join queries. In this paper, we study the problem of enumerating the results of a join query in random order. We develop an efficient random-order enumeration algorithm for join queries with no large hidden constants in its complexity, achieving expected $O(\frac{\mathrm{AGM}(Q)}{|Res(Q)|}\log^2|Q|)$ delay, $O(\mathrm{AGM}(Q)\log|Q|)$ total running time after $O(|Q|\log|Q|)$-time index construction, where $|Q|$ is the size of input, $\mathrm{AGM}(Q)$ is the AGM bound, and $|Res(Q)|$ is the size of the join result. We prove that our algorithm is near-optimal in the worst case, under the combinatorial $k$-clique hypothesis. Our algorithm requires no query-specific preprocessing and can be flexibly adapted to many common database indexes with only minor modifications. We also devise two non-trivial techniques to speed up the enumeration, and provide an experimental study on our enumeration algorithm along with the speed-up techniques. The experimental results show that our algorithm, enhanced with the proposed techniques, significantly outperforms existing state-of-the-art methods.
\end{abstract}

\maketitle

\pagestyle{\vldbpagestyle}
\begingroup\small\noindent\raggedright\textbf{PVLDB Reference Format:}\\
\vldbauthors. \vldbtitle. PVLDB, \vldbvolume(\vldbissue): \vldbpages, \vldbyear.\\
\href{https://doi.org/\vldbdoi}{doi:\vldbdoi}
\endgroup
\begingroup
\renewcommand\thefootnote{}\footnote{\noindent
This work is licensed under the Creative Commons BY-NC-ND 4.0 International License. Visit \url{https://creativecommons.org/licenses/by-nc-nd/4.0/} to view a copy of this license. For any use beyond those covered by this license, obtain permission by emailing \href{mailto:info@vldb.org}{info@vldb.org}. Copyright is held by the owner/author(s). Publication rights licensed to the VLDB Endowment. \\
\raggedright Proceedings of the VLDB Endowment, Vol. \vldbvolume, No. \vldbissue\ %
ISSN 2150-8097. \\
\href{https://doi.org/\vldbdoi}{doi:\vldbdoi} \\
}\addtocounter{footnote}{-1}\endgroup

\ifdefempty{\vldbavailabilityurl}{}{
\vspace{.3cm}
\begingroup\small\noindent\raggedright\textbf{PVLDB Artifact Availability:}\\
The source code, data, and/or other artifacts have been made available at \url{\vldbavailabilityurl}.
\endgroup
}
\section{Introduction}
As one of the most fundamental types of queries in database systems,
join queries\footnote{The join queries discussed in this paper are more precisely known as equi-joins.} combine columns from one or more tables into a new table based on equality conditions.
The efficient evaluation algorithm of join queries has been extensively studied theoretically and practically in the database community.
Many of these algorithms~\cite{WCOJOIN2012,genericjoin,WCOJOIN2014,WCOJOIN2014B,WCOJOIN2016,WCOJOIN2018,WCOJOIN2020,WCOJOIN2021} can achieve asymptotically optimal performance (sometimes up to a polylogarithmic factor) in worst-case conditions.
However, joins remain computationally expensive, even with the implementation of worst-case optimal algorithms.
The major reason for this is the size of the join result:
a join query $Q$ may produce tuples as many as the AGM bound~\cite{AGMBound}, which is usually much larger than the size of the relation tables of $Q$.
Then, even listing all tuples in the join result requires a large amount of time in the worst case.
This poses a significant challenge in database systems and data analysis pipelines~\cite{Sample2023}.


To address the aforementioned challenge,
numerous algorithms have been proposed that generate and output result tuples to downstream tasks in a continuous manner~\cite{Enum2007,Enum2013,Enum2018,Enum2019,Enum2020},
no waiting for all join results to be generated in advance.
These algorithms, say \emph{enumeration algorithms}, offer the advantage that the number of intermediate results is proportional to the processing time.
This is useful when the query is a part of a larger data analysis pipeline, where the query results are sent to downstream processing~\cite{REnumUCQ2022}.
For example, in machine learning pipelines, we may use query results as the training data or the features for the task of training a model~\cite{StreamLearning2001,trainapp2018,trainapp2020}.
In such pipelines, intermediate results can be fed to the next-step process without waiting for complete result materialization. This can be seen in streaming learning algorithms~\cite{StreamLearning2001}. In addition, users are able to halt the training when the learning process reaches a stable state.

Furthermore, enumeration in random order is required, \emph{i.e.}, each query result tuple is enumerated uniformly at random from the set of result tuples not yet enumerated~\cite{REnumUCQ2022}. In this way, at any time, the algorithm outputs a statistically meaningful representation of all the join results.
Many downstream tasks, where the complete query result is not necessarily required, can benefit significantly from the random samples, and the benefit increases with the size of the samples.
To this end, Carmeli et al.~\cite{REnumUCQ2022} introduced a random-order enumeration algorithm for free-connex acyclic join queries with a polylogarithmic delay after a linear-time preprocessing phase.
They also prove that (under complexity assumptions) there is no random-order enumeration algorithm with linear preprocessing and polylogarithmic delay for non-free-connex conjunctive queries,
and argue that it would be interesting to understand more precisely the time it is required (both in terms of enumeration delay and preprocessing time) for enumerating results of non-free-connex queries (such as cyclic join queries) in random order.

For general join queries, a straightforward method~\cite{IncEm2019,Sample2023} is to develop a trivial transformation from a uniform sampling (with replacement) algorithm \cite{Sample2018,Sample2020,pgmjoin,Sample2023} into a sampling without replacement process.
This is typically done by treating the sampling algorithm as a black box that generates candidate results uniformly and discarding any duplicates that have already been seen.
To sample (cyclic) join results,
early works~\cite{Sample2018,pgmjoin} decompose the original cyclic query into an acyclic join query (say the skeleton query) and a residual component.
They then apply an acyclic‑join sampling algorithm on the skeleton query and employ a trivial procedure to sample the join results between the samples of the skeleton query and the residual component.
Although these methods are practically motivated, they do not achieve near-optimal running time, and require at least linear time preprocessing for each query before sampling.
More recent works~\cite{Sample2023, Sample2023A, Tao2024} avoid decomposition and directly perform sampling on cyclic joins.
These algorithms achieve near-optimal expected sampling time under complexity assumptions, without query-specific preprocessing.
However, the constant and polylogarithmic factors of their running time are typically large, which scale exponentially with the query size (\emph{e.g.}, the expected running time of~\cite{Sample2023} is actually \begin{equation}O\left(\frac{\mathrm{AGM}(Q)}{\max\{1,|Res(Q)|\}}\log^{d+1}|Q|\right),
\end{equation}
where $d$ is the maximum arity of the relations and $Res(Q)$ denotes the join result). In addition, to the best of our knowledge, these algorithms have not yet been implemented and practically evaluated.

However, the sampling-based enumeration method wastes a lot of time in practice, as many generated result tuples will be discarded.
There is also no worst-case guarantee on its total running time,
and there is no strict guarantee that all result tuples can be output when the algorithm ends.
Strictly speaking, it would not even be counted as an enumeration algorithm with a delay sublinear to the number of query results~\cite{REnumUCQ2022}.
Worse still, sampling-based random-order enumeration inherits the aforementioned weaknesses of existing join sampling algorithms.

In this paper, we present an efficient random-order enumeration algorithm for join queries that provides a worst-case guarantee on the total running time.
Theoretically, our algorithm enumerates the join result tuples in expected \begin{equation}
    O\left(\frac{\mathrm{AGM}(Q)}{|Res(Q)|+1}\log^2|Q|\right)
\end{equation} delay and 
\begin{equation}
    O(\mathrm{AGM}(Q)\log|Q|)
\end{equation} total running time, after an $O(|Q|\log|Q|)$-time index construction phase. And it is proved to be a nearly worst-case optimal algorithm, and achieves the same total running time complexity with several worst-case optimal join algorithms such as Generic Join~\cite{genericjoin}. 
Practically, we develop techniques to speed up the enumeration, which lead to a significant performance improvement in our experiments, especially when $\mathrm{AGM}(Q)\gg|Res(Q)|$.
Our algorithm can not only serve as an near-optimal random-order enumeration algorithm without large constant or polylogarithmic factor in its enumeration delay and total running time; it can also be used as an efficient join sampling algorithm without preprocessing, \emph{i.e.}, the indexes we build over the relational tables can be reused for different join queries, without the need for linear-time preprocessing for each query.
Moreover, our algorithm does not require complex index structures built for relation tables.
It supports various hierarchical indexing structures that are widely used in database systems,
such as balanced trees, B-trees, skip lists, and tries.
When updates are not required, simply sorting the tuples in lexicographic order suffices as a lightweight indexing mechanism for our algorithm.
This flexibility enables our algorithm to be integrated into different database systems with minimal development overhead.
Concretely, we list our contributions in the following.


First, we develop an efficient random-order enumeration algorithm framework for join queries, based on a novel concept, \textsc{RRAccess} (the Relaxed Random-Access algorithm) and a well-organized data structure (the Ban-Pick tree), which are first proposed in this paper.
Specifically, \textsc{RRAccess} is defined to calculate a bijection $\alpha$ from a set of integers $S$ (not necessarily consecutive) to the set of join result tuples $Res(Q)$,
    and it returns a trivial interval $I$ with $I\cap S=\emptyset$ when inputting an integer not in $S$.
    The Ban-Pick tree maintains a collection of disjoint intervals, referred to as the banned intervals,
    and it supports picking an integer from the remaining (unbanned) intervals uniformly at random.
    Based on the Ban-Pick tree, we show that if there is an \textsc{RRAccess} algorithm that runs in $O(\log^2|Q|)$ worst-case time and $O(\log|Q|)$ amortized time and satisfies $\forall s\in S, s\le \mathrm{AGM}(Q)$,
    then our algorithm enumerates result tuples in $Res(Q)$ with expected $O(\frac{\mathrm{AGM}(Q)}{|Res(Q)|+1}\log^2|Q|)$ delay and $O(\mathrm{AGM}(Q)\log|Q|)$ total running time.
    And it is proved as a nearly worst-case optimal algorithm since its expected enumeration delay and total running time are only polylogarithmically higher than the theoretical lower bound.

    
    Second, we design an \textsc{RRAccess} algorithm based on a logical tree structure referred to as the relaxed random-access tree (RRATree),
    where each node of RRATree corresponds to a filter, and the set of result tuples satisfying a parent node’s filter is recursively partitioned into subsets corresponding to its children’s filters.
    By fully leveraging the properties of filters in RRATree, we construct efficient data structures and develop an upper-bound estimation algorithm and a children exploration algorithm, both of them run in $O(\log|Q|)$ time. These components enable \textsc{RRAccess} to achieve the $O(\log^2|Q|)$ worst-case time and $O(\log|Q|)$ amortized time.
    Thus yielding the nearly worst-case optimal \textsc{REnum}.
    
    
    Third, based on our algorithm framework, two non-trivial techniques are developed to speed up the enumeration.
    The first is \emph{larger trivial interval discovery} (LTI).
    Observe that the trivial integers in $\mathbb{N}[1,\mathrm{AGM}(Q)]\setminus\alpha^{-1}(Res(Q))$ are often continuous and can be grouped into intervals (called trivial intervals).
    LTI discovers large trivial intervals during the running of \textsc{RRAccess} to avoid picking more trivial integers in the follow-up steps.
    Consequently, the probability that \textsc{RRAccess} returns \textit{false} is reduced, so that the total running time and enumeration delay are reduced.
    To further reduce the probability that \textsc{RRAccess} returns \textit{false}, we develop the second technique named \emph{tighter upper-bound estimation} (TU).
    TU provides a tighter upper-bound estimation of the number of the filtered join results for each filter in the RRATree,
    so that more and larger trivial intervals are discovered and banned earlier.
    As a result, it improves the effectiveness of LTI.
    
    
At last, we validate the efficiency of our algorithm along with the speed-up techniques through experiments.
The experiment results show that with these techniques, our algorithm significantly outperforms the sampling-based algorithms~\cite{pgmjoin,Sample2023}.

The remainder of this paper is structured as follows.
The basic notations and some necessary techniques are introduced in Section~\ref{SEC:PRELIMINARIES}.
In Section~\ref{SEC:OVERVIEW} we introduce our main random-order enumeration algorithm framework.
Section~\ref{SEC:RRATree} details the implementation of the \textsc{RRAccess} algorithm and RRATree.
Section~\ref{SEC:OPTIMIZE} discusses non-trivial speed-up techniques that significantly speed up the enumeration in practice.
And Section~\ref{SEC:EXPERIMENTS} presents our experimental study.

\section{Preliminaries}\label{SEC:PRELIMINARIES}

In this paper, we denote the set of natural numbers by $\mathbb{N}$. For any natural numbers $i$ and $j$ with $i \leq j$, we define $\mathbb{N}[i, j]=[i,j]\cap\mathbb{N}$.

\subsection{Join Query} Given a finite attribute set $\mathrm{Att}$ and a set $U\subseteq \mathrm{Att}$, a \emph{tuple} over $U$ is a function $t:U\rightarrow\mathbb{N}$, where $\mathbb{N}$ is the set of all integers, and a projection of a tuple $t$ on $V\subseteq U$, say $t[V]$, is a tuple that satisfies $t[V](v)=t(v)$ for each $v\in V$. A relation $R$ is a set of tuples over an identical attribute set $U$, say $\mathrm{att}(R)=U$.
Then a join query $Q$ is defined as a set of relations $\{R_1, \dots, R_m\}$, and can be expressed as $Q=R_1\Join\dots\Join R_m$. Then let $|Q|=\sum_{i=1}^m|R_i|$ be the sum of sizes of the relations in $Q$ (\emph{i.e.} the size of input), and the result of the query $Q$ is defined as follows:
\begin{equation}
\mathit{Res}(Q)=\{t\text{ over }\mathrm{att}(Q)|\forall R\in Q: t[\mathrm{att}(R)]\in R\},
\end{equation}
where $\mathrm{att}(Q)=\bigcup_{R\in Q}\mathrm{att}(R)$.
Let $\textrm{dom}_Q(v)$ denote the \emph{active domain} of the attribute $v$, more specifically,
\begin{equation}
\textrm{dom}_Q(v)=\bigcup_{R\in Q}\bigcup_{v\in\mathrm{att}(R)}\{t(v)|t\in R\},
\end{equation}
then we have $\mathit{Res}(Q)\subseteq\prod_{v\in\mathrm{att}(Q)}\mathrm{dom}_Q(v)$.

\subsection{AGM Bound} Given a join query $Q$, the \emph{schema graph} of $Q$ is defined as a hypergraph $G_Q=(V,E)$, where $V=\mathrm{att}(Q)$ and $E=\{\mathrm{att}(R)|R\in Q\}$. Let $c:E\rightarrow(0,1)$ be a fractional edge cover of $G_Q$, \emph{i.e.}, $\forall v\in V$,
$\sum_{v\in e}c(e)\ge 1$.
Then $|\mathit{Res}(Q)|\le \textrm{AGM}_c(Q)$~\cite{AGMBound}, where
\begin{equation}
\textrm{AGM}_c(Q):=\prod_{R\in Q}|R|^{c(\mathrm{att}(R))}.
\end{equation}

Moreover, let $EC(G_Q)$ denotes the set of all fractional edge covers of $G_Q$, the minimized AGM bound of $Q$, say $\textrm{AGM}(Q)=\min_{c\in EC(G_Q)}\textrm{AGM}_c(Q)$, is tight: there exists a join query $Q^*$ that satisfies $|\mathit{Res}(Q^*)|=\Omega(\textrm{AGM}(Q^*))$~\cite{AGMBound}.
Note that the AGM bound can be efficiently computed with only the sizes of the relation tables, and the minimized AGM bound can be calculated by solving a linear program in $O(1)$ time under data complexity.

\subsection{Uniform Sampling} 
A \emph{join sampling algorithm} outputs each join result tuple with an identical probability.    
Formally, a join sampling algorithm is a randomized algorithm $\mathcal{G}$ that takes a join query $Q$ as input and outputs a tuple in $Res(Q)$, such that
\begin{equation}
    \forall t\in \mathit{Res}(Q), \mathrm{Pr}(\mathcal{G}(Q)\text{ outputs }t)=\frac{1}{|\mathit{Res}(Q)|}.
\end{equation}



By Deng et al.~\cite{Sample2023}, there is a nearly worst-case optimal join sampling algorithm for join queries (under a complexity hypothesis):
\begin{theorem}[\cite{Sample2023}]\label{thm:WCOLVUS}
    There is a uniform join sampling algorithm running in expected $\Tilde{O}(\frac{\mathrm{AGM}(Q)}{\max\{1,|Res(Q)|\}})$ time after a $\Tilde{O}(|Q|)$-time index construction phase.
    Moreover, under the combinatorial $k$-clique hypothesis, for any $\varepsilon>0$, there is no uniform sampling algorithm for join queries that runs in $\Tilde{O}(|Q|+\frac{|Q|^{{\rho^*}-\varepsilon}}{|\mathit{Res}(Q)|})$ time with high probability, where $\rho^*$ is the fractional edge cover number of $G_Q$.
\end{theorem}

\subsection{Random-Order Enumeration}
The random-order enumeration algorithm for join queries is a random algorithm that takes a join query $Q$ as input and outputs all tuples in $\mathit{Res}(Q)$ in random order.
In other words, for each $1\le i<|\mathit{Res}(Q)|$, after the first $i$ result tuples $t_1,\dots,t_i$ have been output, the $(i+1)$-th output result tuple can be seen as a tuple uniformly sampled from $\mathit{Res}(Q)\setminus \{t_1,\dots,t_i\}$.
The (expected) enumeration delay of a random-order enumeration algorithm is defined as the maximum (expected) length of the following time intervals: (1) the time from the start of the algorithm to the output of first result tuple, (2) the time from the output of any result tuple to the output of the next result tuple, and (3) the time from the output of the last result tuple to the terminate of the algorithm.

\section{Enumeration Framework Overview}\label{SEC:OVERVIEW}

In this section, we develop an efficient random-order enumeration algorithm framework with a worst-case guarantee on its total running time.
To facilitate an intuitive exposition of our algorithm,
we begin by defining a representative cyclic join query, denoted $Q_\Delta$, which will serve as a running example in the subsequent discussion.

\begin{table}[]
\begin{subfigure}[b]{0.3\linewidth}
\centering
\begin{tikzpicture}[scale=0.25]
    \coordinate (y) at (0,0);
    \coordinate (z) at (4,0);
    \coordinate (x) at (2,3);

    \filldraw[black] (x) node {$x$};
    \filldraw[black] (y) node {$y$};
    \filldraw[black] (z) node {$z$};

    \draw (2,0) ellipse (3.5 and 1) node[below=0.4cm]{$S$};
    \draw[rotate around={56.31:(1,1.5)}] (1,1.5) ellipse (3.5 and 1) node[above left=0.2cm]{$R$};
    \draw[rotate around={-56.31:(3,1.5)}] (3,1.5) ellipse (3.5 and 1) node[above right=0.2cm]{$T$};

\end{tikzpicture}
        \caption{$G_{Q_\Delta}$}
\end{subfigure}
        \hfill
    \begin{minipage}[b]{0.69\linewidth}
    \centering
    \begin{subtable}[b]{0.2\linewidth} 
        \centering
        \begin{tabular}{cc}
        \toprule
        $x$ & $y$ \\
        \midrule
        1 & 2 \\
        2 & 3 \\
        3 & 4 \\
        4 & 1 \\
        \bottomrule
        \end{tabular}
        \caption{$R$}
        \label{tab:R}
    \end{subtable}
    \hfill 
    \begin{subtable}[b]{0.2\linewidth} 
        \centering
        \begin{tabular}{cc}
        \toprule
        $y$ & $z$ \\
        \midrule
        1 & 3 \\
        3 & 4 \\
        4 & 4 \\
        4 & 1 \\
        \bottomrule
        \end{tabular}
        \caption{$S$}
        \label{tab:S}
    \end{subtable}
    \hfill 
    \begin{subtable}[b]{0.2\linewidth} 
        \centering
        \begin{tabular}{cc}
        \toprule
        $x$ & $z$ \\
        \midrule
        2 & 4 \\
        3 & 1 \\
        3 & 4 \\
        4 & 2  \\
        \bottomrule
        \end{tabular}
        \caption{$T$}
        \label{tab:T}
    \end{subtable}
    \hfill 
    \begin{subtable}[b]{0.3\linewidth} 
        \centering
        \begin{tabular}{ccc}
        \toprule
        $x$ & $y$ & $z$ \\
        \midrule
        2 & 3 & 4 \\
        3 & 4 & 1\\
        3 & 4 & 4 \\
        \bottomrule\\
        \end{tabular}
        \caption{$Res(Q_\Delta)$}
        \label{tab:RES}
    \end{subtable}
    \end{minipage}
    \caption{$Q_\Delta:=R\Join S\Join T$}
    \label{tab:all}
\end{table}

\begin{example}[$Q_\Delta$]\label{example}
Let $Q_{\triangle}:=R\Join S\Join T$, in which $\mathrm{att}(R)=\{x,y\},\mathrm{att}(S)=\{y,z\}$ and $\mathrm{att}(T)=\{x,z\}$.
The schema graph, tables and join results of $Q_\Delta$ are shown in Table~\ref{tab:all}.
\end{example}

To enumerate the result tuples of $Q_\Delta$ in uniformly random order,
one can construct a bijective mapping between the integer range $\mathbb{N}[1,|\mathit{Res}(Q_\Delta)|]$ and the set of result tuples.
For example,
sort $\mathit{Res}(Q_\Delta)$ in lexicographical order and let each integer $i\in\mathbb{N}[1,|\mathit{Res}(Q_\Delta)|]$ maps to the $i$-th tuple, say $\mathit{Res}(Q_\Delta)^\pi(i)$.
By generating a random permutation of these integers, one obtains a random sequence $i_1,\dots,i_N$, then outputting $Res(Q_\Delta)^\pi(i_k)$ for $k=1,\dots,N$ yields a random-order enumeration of $\mathrm{Res}(Q_\Delta)$.
In Example~\ref{example}, if the order of the random permutation of integers is $2, 3, 1$, then the output of the random enumeration algorithm is $(3,4,1),(3,4,4),(2,3,4)$.
However, listing and sorting the join results would incur prohibitively high preprocessing time.
In fact, for cyclic join queries, no such bijection can be computed efficiently (\emph{i.e.}, in polylogarithmic time after linear time preprocessing), under some complexity hypotheses~\cite{REnumUCQ2022}.  

In our approach, we relax the requirement of a bijective mapping from a sequence of consecutive integers to the result tuples.
This relaxation implies that not every integer within the sequence must correspond to a result tuple, some integers,
referred to as \emph{trivial integers}, are instead mapped to ``false", indicating that they do not correspond to any join result.
The remaining integers, called \emph{non-trivial integers}, are each mapped to a result tuple,
that is, the mapping between the non-trivial integers and the result tuples remains bijective.
Then we pick integers uniformly at random without replacement and output the result tuple they correspond to (if exist).
Obviously, this process forms a random-order enumeration of the join result.
Moreover, to avoid picking trivial integers as much as possible,
if a trivial integer is picked, we discover a \emph{trivial interval} consisting of trivial integers,
and design an efficient data structure to maintain the set of discovered trivial intervals.
Once a trivial interval is discovered, the integers in it will not be picked in the follow-up steps.

The remainder of this section will introduce (1) the formal definition of the mapping between integers and join results, and an algorithm to calculate it (\emph{i.e.,} the relaxed random-access algorithm), (2) a novel data structure that supports online banning of intervals and uniform sampling of unbanned integers (\emph{i.e.,} the Ban-Pick tree), and (3) an efficient random-order enumeration algorithm framework based on (1) and (2).

\subsection{The Relaxed Random-Access Algorithm}
Given a family of functions $\varphi=\left\{\varphi_Q|Q\in\mathcal{Q}\right\}$ and a join query $Q$, $\varphi_Q:\mathbb{N}^+\rightarrow \mathit{Res}(Q)\cup\{\textit{false}\}$ satisfies that for each tuple $t\in \mathit{Res}(Q)$ there exists only one $i\in \mathbb{N}^+,\varphi_Q(i)=t$.
Let $N$ be an upper-bound of $|\mathit{Res}(Q)|$ satisfying $\forall i>N,\varphi_Q(i)=\textit{false}$.
We observe that the integers in $\{i|\varphi^*(i)=\textit{false}\}$, say the \emph{trivial integers}, are often continuous and can be grouped into intervals (say the \emph{trivial intervals}), especially when $\mathrm{AGM}(Q)\gg|Res(Q)|$.
In order to prevent from picking as many trivial integers as possible, the relaxed random‐access algorithm reports the trivial intervals.
Formally, we define $\textsc{RRAccess}^{Q,\varphi}$ as the algorithm that takes an integer $i$ as input and behaves as follows:
\begin{enumerate}
    \item returns $\varphi_Q(i)$ if $\varphi_Q(i)\neq false$,
    \item otherwise, returns a trivial interval $[a,b]\subseteq \mathbb{N}^+$ (with $a\le i\le b$) containing only trivial integers.
\end{enumerate}

We will introduce the implementation of the relaxed random-access algorithm that runs in $O(\log^2|Q|)$ worst-case time and $O(\log|Q|)$ amortized time in Section~\ref{SEC:RRATree}.

\subsection{The Ban-Pick Tree} 
To avoid repeatedly picking integers that have already been picked or that are known to not correspond to any join result,
we dynamically maintain a collection of intervals representing such integers, which are excluded from further picking.
Specifically, we define two types of integers:  
(1) \emph{trivial integers}, which do not correspond to any join result, and  
(2) \emph{picked integers}, which have already been picked during previous steps.
We maintain a set $B$ of disjoint intervals, each consisting of either trivial or picked integers, using a data structure called the \emph{Ban-Pick tree}.  
This structure, together with two operations, guarantees that newly picked integers do not fall within any interval in $B$.
Formally, the Ban-Pick tree enables the following two operations:
\begin{enumerate}
    \item \textbf{the ban operation $\bm{B.ban}$}, that takes an interval disjoint from all intervals in $B$ as input and inserts it into $B$,
    \item \textbf{the pick operation $\bm{B.pick}$}, that takes an integer $H$ satisfying $\forall I\in B, I\subseteq [1,H]$ as input and returns an integer $i\in [1,H]\setminus \cup_{I\in B}I$ uniformly at random.
\end{enumerate}
Based on the Ban-Pick tree, the trivial and picked integers in the banned intervals in $B$ are prevented from being picked.

In general, the set $\mathbb{N}[1,N] \setminus \bigcup_{I\in B} I$ is not a contiguous interval.
This results in a failure of the trivial generator working on an interval.
Therefore, we need to provide a pick operation working on a union of disjoint intervals.
Given $B=\{I_i=[l_i,h_i]|i\in\mathbb{N}[1,|B|]\}$ which is the set of disjoint intervals already banned,
\emph{w.l.o.g.}, assume $h_{i}<l_{i+1}$ for every $i\in\mathbb{N}[1,|B|-1]$.
Let $L=\left|\mathbb{N}[1,N]\setminus \cup_{i=1}^{|B|}I_i\right|$, that is, $L=N-\sum_{i=1}^{|B|}{|I_i|}$.
Our pick operation works as follows:
\begin{enumerate}
    \item sample an integer $y\in\mathbb{N}[1,L]$ uniformly at random,
    \item compute the offset $b=\sum_{i=1}^{k^*}|I_i|$ such that $h_{k^*}<{y+b}$ and $y+b<l_{k^*+1}$ (if $k^*<|B|$),
    \item return $y+b$.
\end{enumerate}
To enable efficient computation of the offset in step (2), we define the Ban-Pick tree as a balanced tree $T_B$ such that


\begin{enumerate}
    \item each node $u\in T_B$ corresponds bijectively to an interval $I_u = [u.l, u.h] \in B$, and stores it via $u.l$ and $u.h$,
    \item each node $u\in T_B$ maintains $u.\textit{left}$ and $u.\textit{right}$ that point to its left child and right child,
    and if $v$ is the left (right) child of $u$, then $v.h < u.l$ ($v.l > u.h$),
    \item each node $u\in T_B$ maintains $u.\textit{take}$, which denotes the sum of lengths of the intervals in the subtree rooted in $u$,
    \item the height of $T_B$ is $O(\log |B|)$.
\end{enumerate}
Obviously, for the ban operation, the time to maintain $T_B$ is $O(\log|B|)$ for each interval insertion.



 \begin{algorithm}[!htbp]
    \caption{$B.pick$}\label{ALG:G}
    \LinesNumbered
    \KwIn{$H$}
    \KwOut{a uniform sample from $\mathbb{N}[1,H]\setminus\cup_{I\in B}I$}
    $u\leftarrow$ the root of $T_B$\;
    pick an integer $y\in\mathbb{N}[1,H-u.take]$ randomly and uniformly\;
    $b\leftarrow0$, $\textit{temp}\leftarrow0$\;
    \While{$u\neq{\textit{nil}}$}{
        \lIf{$u.\textit{left}=\textit{nil}$}{$\textit{temp}\leftarrow0$}
        \lElse{$\textit{temp}\leftarrow u.\textit{left}.\textit{take}$}
        \lIf{$(y+b)+\textit{temp}<u.l$}{$u\leftarrow u.\textit{left}$}
        \lElse{$b\leftarrow b+\textit{temp}+(u.h-u.l+1)$, $u\leftarrow u.\textit{right}$}
    }
    \Return{$y+b$}
\end{algorithm}

For the pick operation, we provide an efficient implementation based on the Ban-Pick tree as Algorithm~\ref{ALG:G}. 
It begins by sampling a random integer $y$ uniformly from the interval $([1, H - u.take])$.
This interval corresponds to the concatenation of all ``gaps'' (unbanned intervals in $[1,H]$).
Then it traverses the tree $T_B$ starting from the root, aiming to locate the gap corresponding to $y$ and calculate the offset $b$.
Finally, it returns the value $y + b$, which lies in $\mathbb{N}[1,H]\setminus\cup_{I \in B} I$.
The height of $T_B$ is at most $O(\log |B|)$; hence, Algorithm~\ref{ALG:G} runs in $O(\log |B|)$ time.
Moreover, since $y$ is sampled from the concatenation of all gaps, the probability of $y$ corresponding to a particular gap is proportional to its size,
and $y+b$ is uniformly distributed within the gap associated with $y$.
Therefore, Algorithm~\ref{ALG:G} produces a uniformly random sample from the set $\mathbb{N}[1, H]\setminus \bigcup_{I \in B} I$.



\subsection{The Enumeration Framework}
Given any join query $Q$, let $N$ be an upper bound of $|\mathit{Res}(Q)|$ such that $\forall i>N,\varphi_Q(i) = \textit{false}$.  
Algorithm~\ref{ALG:BPREnum} enumerates all tuples in $\mathit{Res}(Q)$ by repeatedly picking integers uniformly at random from $[1, N] \setminus B$, where $B$ is a set of banned intervals containing trivial and picked integers.  
Each picked integer $i$ is passed to $\textsc{RRAccess}^{Q,\varphi}$.
If $\varphi_Q(i)\neq false$, in which case $\textsc{RRAccess}^{Q,\varphi}$ returns a valid result tuple, then the algorithm outputs the result tuple, and $[i,i]$ is inserted into $B$ to avoid repetition;  
otherwise, if $\varphi_Q(i) = \textit{false}$, indicating that $i$ lies in a trivial interval $I$ returned by $\textsc{RRAccess}^{Q,\varphi}$, then the entire interval $I$ is inserted into $B$.  
This process ensures that each result tuple is output exactly once.
Moreover, at each step, the output tuple (if exists) is uniformly sampled from the set of unenumerated join results.

\begin{algorithm}
    \caption{\textsc{REnum}}\label{ALG:BPREnum}
    \LinesNumbered
    \KwIn{$Q$}
    \KwOut{$\mathit{Res}(Q)$ in a random order}
    $B\leftarrow\emptyset$, $N_\mathrm{ban}\leftarrow 0$\;
    \While{$N_\mathrm{ban}<N$}{
        $i\leftarrow B.pick(N)$\;
        $res\leftarrow\textsc{RRAccess}^{Q,\varphi^*}(i)$\;
        \If{$res$ is a tuple in $\mathit{Res}(Q)$}{\textbf{output} $res$\; $B.ban([i,i])$, $N_\mathrm{ban}\leftarrow N_\mathrm{ban} + 1$;}
        \Else(\hspace{1ex}//\texttt{\footnotesize  $res$ is a trivial interval.}){$B.ban(res)$, $N_\mathrm{ban}\leftarrow N_\mathrm{ban} + |res|$;}
    }
\end{algorithm}

\begin{lemma}\label{lem:RREnum}
    For any $Q\in\mathcal{Q}$, if $\log N\le O(\log |Q|)$ and there exists a relaxed random-access algorithm $RRAccess^{Q,\varphi}$ running in $O(\log^2|Q|)$ worst-case time and $O(\log|Q|)$ amortized time, then Algorithm~\ref{ALG:BPREnum} enumerates the result tuples in $\mathit{Res}(Q)$ in random order in expected $O(\frac{N}{|\mathit{Res}(Q)|+1}\log^2|Q|)$ delay.
    And its total running time is at most $O(N\log|Q|)$.
\end{lemma}

\begin{proof}
    \PY{
    To simplify the analysis, we assume that $\textsc{RRAccess}^{Q,\varphi}(i)$ returns the interval $[i, i]$ when $\varphi_Q(i) = false$.
    This choice of a minimal trivial interval, while less efficient, enables us to derive upper bounds on both the total running time and the delay of Algorithm~\ref{ALG:BPREnum}.

    \emph{Total running time.} Since $\log|B|\le \log N\le O(\log|Q|)$, all of $B.ban$, $B.pick$ and \textsc{RRAccess} run in $O(\log |Q|)$ amortized time.
    Moreover, since they run at most $N$ times, the total running time of Algorithm~\ref{ALG:BPREnum} is at most $O(N\log|Q|)$.
    }

    \emph{Enumeration delay.} Let $N_i(0\le i\le |\mathit{Res}(Q)|)$ be the random variable of the number of times that $\textsc{RRAccess}^{Q,\varphi}$ returns \textit{false} in the process of outputting the first $i$ result tuples (the $(i+N_i)$-th run of $\textsc{RRAccess}^{Q,\varphi}$ returns the $i$-th result tuple, and in particular, $N_0$=0).
    By the definition of $\varphi_Q$, for each tuple $t\in \mathit{Res}(Q)$,
    there exists only one integer $i \in \mathbb{N}[1, N]$ such that $\varphi_Q(i) = t$.
    Then there are exactly $|\mathit{Res}(Q)|$ non-trivial integers in $\mathbb{N}[1, N]$,
    and for each $0\le i\le|\mathit{Res}(Q)|$, $N_i$ is the number of trivial integers obtained by sampling (without replacement) from $\mathbb{N}[1, N]$ until exactly $i$ non-trivial integers are picked.
    This implies that for each $1\le i\le|\mathit{Res}(Q)|$, $N_i$ follows the negative hypergeometric distribution~\cite{neghyperdistribution} with parameters $N$, $N-|\mathit{Res}(Q)|$ and $i$, and its expectation is:
    \begin{equation}
        E[N_i]=\frac{i\left(N-|\mathit{Res}(Q)|\right)}{|\mathit{Res}(Q)|+1}.
    \end{equation}
    
    Therefore, the expected number of calls of $\textsc{RRAccess}^{Q,\varphi}$ between the output of the $(i-1)$-th result (the start of the algorithm when $i=1$) and the output of the $i$-th result is 
    \begin{equation}
            E\left[N_i-N_{i-1}+1\right]=\frac{N+1}{|\mathit{Res}(Q)|+1}=O\left(\frac{N}{|\mathit{Res}(Q)|+1}\right).
    \end{equation}
    And the expected running time of $\textsc{RRAccess}^{Q,\varphi}$ between the output of the last result tuple and the end of the algorithm is
    \begin{equation}
    \begin{aligned}
        E\left[N-N_{|\mathit{Res}(Q)|}-|\mathit{Res}(Q)|\right]&= \frac{N-|\mathit{Res}(Q)|}{|\mathit{Res}(Q)|+1}=O\left(\frac{N}{|\mathit{Res}(Q)|+1}\right).
    \end{aligned}
    \end{equation}
    Then, since $\textsc{RRAccess}^{Q,\varphi}$ runs in $O(\log^2|Q|)$ worst-case time,
    the expected delay of Algorithm~\ref{ALG:BPREnum} is $O(\frac{N}{|\mathit{Res}(Q)|+1}\log^2|Q|)$.    
\end{proof}

In the case where $N\le \mathrm{AGM}(Q)$, in other words, if $\textsc{RRAccess}^{Q,\varphi}$ in Lemma~\ref{lem:RREnum} satisfies $\forall i>\mathrm{AGM}(Q), \varphi(i)=\textit{false}$, then Algorithm~\ref{ALG:BPREnum} enumerates the join results in expected $O(\frac{\mathrm{AGM}(Q)}{|\mathit{Res}(Q)|+1}\log^2|Q|)$ delay and $O(\mathrm{AGM}(Q)\log|Q|)$ total running time, where $\rho^*$ is the fractional edge cover number of $G_Q$.

In addition, by \cite{Sample2023}, the $O(\mathrm{AGM}(Q)\log|Q|)$ total running time is nearly optimal in the worst case, unless the combinatorial $k$-clique hypothesis is false.
Moreover, since the first enumerated result tuple is a uniform sample from $Res(Q)$,
the lower bound of the expected running time of join sampling algorithms is also a lower bound of the expected delay of random-order enumeration algorithms.
Formally, the following theorem holds.
\begin{theorem}\label{thmm}
    Under the combinatorial $k$-clique hypothesis, for any $\varepsilon>0$, there is no random-order enumeration algorithm for join queries with expected $\Tilde{O}(\frac{|Q|^{\rho^*-\varepsilon}}{|\mathit{Res}(Q)|+1})$ delay after a $\Tilde{O}(|Q|)$-time preprocessing, where $\rho^*$ is the fractional edge cover number of $G_Q$.
\end{theorem}
\begin{proof}
    For $\forall\varepsilon>0$,
    if there exists a random-order enumeration algorithm $\mathcal{A}$ for join queries with expected $\Tilde{O}(\frac{|Q|^{\rho^*-\varepsilon}}{|\mathit{Res}(Q)|+1})$ delay after a $\Tilde{O}(|Q|)$-time preprocessing,
    we can define $\mathcal{B}$ as the random algorithm obtained by executing $\mathcal{A}$ until the first result tuple is outputted and then stopping the enumeration.
    Then $\mathcal{B}$ is a uniform sampling algorithm for join queries,
    and according to Markov's inequality, $\mathcal{B}$ runs in $\Tilde{O}(|Q|+\frac{|Q|^{\rho^*-\varepsilon}}{|\mathit{Res}(Q)|})$ time with high probability.
    This contradicts Theorem~\ref{thm:WCOLVUS}.
\end{proof}

This shows that, assuming a preprocessing phase of $\Tilde{O}(|Q|)$ time (\emph{e.g.}, for index construction), the expected $O(\frac{\mathrm{AGM}(Q)}{|\mathit{Res}(Q)|+1}\log^2|Q|)\le O(\frac{|Q|^{\rho^*}}{|\mathit{Res}(Q)|+1}\log^2|Q|)$ delay is nearly optimal.

\section{The Relaxed Random-Access}\label{SEC:RRATree}

In this section, we describe an efficient implementation of \textsc{RRAccess} that runs in $O(\log^2|Q|)$ worst-case time and $O(\log|Q|)$ amortized time.
It is based on a logical tree structure, referred to as the relaxed random-access tree (RRATree), denoted by $\Tilde{T}_Q$.


\subsection{Overview of RRATree and \textsc{RRAccess}}
Formally, the RRATree is defined as follows:

\begin{definition}[RRATree]\label{DEF:RRATREE}

Let $Q$ be an arbitrary join query. The RRATree of $Q$, denoted $\Tilde{T}_Q$, is a rooted tree that satisfies:

\begin{enumerate}
    \item each node $u\in\Tilde{T}_Q$ corresponds to a filter $\psi_u$,
    \item let $r$ be the root node of $\Tilde{T}_Q$, then the filter $\psi_r$ can be computed in $O(|Q|)$ time and satisfies $Q=Q|_{\psi_r}$,
    \item there is an upper-bound algorithm $upp$ which takes a filter $\psi_u$ with $u\in\Tilde{T}_Q$ as input and returns an positive integer in $O(\log|Q|)$ time, such that
    \begin{enumerate}
        \item $\forall u\in\Tilde{T}_Q, upp(\psi_u)\ge |\mathit{Res}(Q|_{\psi_{u}})|$,
        \item for any filter $\psi_u$ with $u\in\Tilde{T}_Q$ and $upp(\psi_u)\le 1$, $\mathit{Res}(Q|_{\psi_u})$ can be computed in $O(\log|Q|)$ time,
    \end{enumerate}
    \item there is a children exploration algorithm $children$, such that for each node $u\in\Tilde{T}_Q$, $children(\psi_u)$ returns at most a constant number of filters $\psi_{v_1},\dots,\psi_{v_k}$ in $O(\log|Q|)$ time, such that
    \begin{enumerate}
        \item $\forall u\in\Tilde{T}_Q$, $children(\psi_u)=\emptyset$ iff $upp(\psi_u)\le 1$,
        \item $\bigcup_{i=1}^k\mathit{Res}(Q|_{\psi_{v_i}})=\mathit{Res}(Q|_{\psi_u})$,
        \item $\forall 1\le i<j\le k,\mathit{Res}(Q|_{\psi_{v_i}})\cap \mathit{Res}(Q|_{\psi_{v_j}})=\emptyset$,
        \item $\sum_{i=1}^k upp(\psi_{v_i})\le upp(\psi_u)$,
        \item $\forall 1\le i\le k, upp(\psi_{v_i})\le\frac{1}{2}upp(\psi_u)$.
    \end{enumerate}
\end{enumerate}

\end{definition}
 
Based on RRATree, we define the family of functions $\varphi^*$ with $upp$ and develop the algorithm \textsc{RRAccess}. Given a join query $Q$ and fix a partial order relation $\prec$ in the filter set, for each integer $i>0$. If there is a tuple $t\in \mathit{Res}(Q)$ such that
\begin{equation}
\label{EQ:phii}
i=\sum_{j=1}^{h-1}\sum_{\substack{\psi\prec \psi_{u_{j+1}}\\\psi\in children(\psi_{u_j})}}upp(\psi)+1,    
\end{equation}
where $u_1,\dots,u_h$ is the path from the root $r$ to the leaf $u_t$ ($u_1=r$ and $u_h=u_t$), and $u_t\in\Tilde{T}_Q$ is the leaf node such that $t\in \mathit{Res}(Q|_{\psi_{u_t}})$,
then $\varphi^*_Q(i)=s$.
Otherwise, if there does not exist such a tuple, we define $\varphi^*_Q(i)=\textit{false}$.
Then the following lemma holds.

\begin{lemma}\label{lem:upperbound}
    $\forall i>upp(\psi_r)$, $\varphi^*_Q(i)=\textit{false}$.
\end{lemma}
\ifshowproof
\begin{proof}
We prove the contrapositive: if $\varphi^*_Q(i) \neq\mathit{false}$, then $i\le upp(\psi_r)$.
    By the definition of $\varphi^*$, for each integer $i>0$ satisfying $\varphi^*_Q(i)\neq \textit{false}$, there is a tuple $t\in \mathit{Res}(Q)$ along with a root-to-leaf path $u_1,\dots,u_h$ ($u_1=r,u_h=u_t$) such that (\ref{EQ:phii}) holds. Then, let
    \begin{equation}
    \begin{aligned}
    \mathrm{sum}_j&=\sum_{\substack{\psi\prec \psi_{u_{j+1}}\\\psi\in children(\psi_{u_j})}}upp(\psi)\\
    &\le \sum_{\psi\in children(\psi_{u_j})}upp(\psi)-upp(\psi_{u_{j+1}})\\
    &\le upp(\psi_{u_j})-upp(\psi_{u_{j+1}}),
    \end{aligned}
    \end{equation}
    therefore
    \begin{equation}
    \begin{aligned}
    i&=\sum_{j=1}^{h-1}\mathrm{sum}_j+1\le \sum_{j=1}^{h-1}\left(upp(\psi_{u_j})-upp(\psi_{u_{j+1}})\right)+1\\
    &=upp(\psi_{u_1})-upp(\psi_{u_{h}})+1=upp(\psi_r).
    \end{aligned}
    \end{equation}
\end{proof}
\fi
\begin{algorithm}[t]
	\caption{$\textsc{RRAccess}^{Q,\varphi^*}$}
	\label{ALG:RRAccess}
        \LinesNumbered
	\KwIn{$i$}
	\KwOut{$\varphi^*_Q(i)$ if $\varphi^*_Q(i)\neq false$, otherwise a trivial interval containing $i$}
        $\mathit{offset}\leftarrow 0$\;
        $\psi\leftarrow \psi_r$ where $r$ is the root node of $\Tilde{T}_Q$\;
	\While{$upp(\psi)\ge 2$}{
            $\psi_1,\dots,\psi_k\leftarrow children(\psi)$\;
            \lIf{$\mathit{offset}+\sum_{j=1}^k upp(\psi_k)<i$}{\textbf{return} $[i,i]$}            
            $r^*\leftarrow \min\{r\in\mathbb{N}[1,k]|\mathit{offset}+\sum_{j=1}^rupp(\psi_j)\ge i\}$\;
            $\mathit{offset}\leftarrow \mathit{offset}+\sum_{j=1}^{r^*-1}upp(\psi_j)$\;
            $\psi\leftarrow\psi_{r^*}$\;
	}
        \If{$\mathit{Res}(Q|_\psi)\neq\emptyset$}{\textbf{return} the tuple in $\mathit{Res}(Q|_\psi)$}
        \lElse{\textbf{return} $[i,i]$}
\end{algorithm}

Our relaxed random-access algorithm is implemented as Algorithm~\ref{ALG:RRAccess}.
The algorithm starts from $\psi_r$, which can be computed in $O(1)$ time after an $O(|Q|\log|Q|)$-time index construction phase.
Then in lines 3-8, the algorithm traverses $\Tilde{T}_Q$ from top to down to find the path on which the filters may contain $\varphi^* _Q(i)$.
Since for each filter $\psi$ and its children $children(\psi)=\psi_1,\dots,\psi_k$, we have $upp(\psi_{i})\le \frac{1}{2}upp(\psi)$ for any $1\le i\le k$,
which implies that the depth of $\Tilde{T}_Q$ is at most $O(\log upp(\psi_r))$ and the number of nodes in $\Tilde{T}_Q$ is $O(upp(\psi_r))$.
Moreover, with the use of a caching mechanism that stores the upper bounds and children of all traversed nodes in $\Tilde{T}_Q$ (thus avoiding redundant computation), in the case where $upp(\psi_r) \le \mathrm{AGM}(Q)$, the following lemma holds.

\begin{lemma}\label{LEM:RRAccess}
If $\psi_r$ can be computed in $O(1)$ time, and there exist an algorithm $upp$ and an algorithm $children$ satisfying the properties in Definition~\ref{DEF:RRATREE} and running in $O(\log|Q|)$ time, then if $upp(\psi_r)\le \mathrm{AGM}(Q)$, Algorithm~\ref{ALG:RRAccess} is a relaxed random-access algorithm running in $O(\log^2|Q|)$ worst-case time and $O(\log|Q|)$ amortized time.
\end{lemma}

\emph{Remarks.} For simplicity, the \textsc{RRAccess} algorithm in this section only implements the trivial interval discovery of a na\"ive method.
Specifically, when $\varphi_Q(i)=false$, the algorithm returns the singleton interval $[i, i]$. 
In Section~\ref{SEC:OPTIMIZE}, we will introduce techniques for discovering larger trivial intervals within the running of \textsc{RRAccess}, with the aim of improving the efficiency of enumeration in practice.


In the following, we introduce the proper filters of the tree nodes, the $O(\log|Q|)$-time upper-bound algorithm and the $O(\log|Q|)$-time children exploration algorithm.

\subsection{The Filters of the Tree Nodes}
In this paper, all filters corresponding to the nodes in $\Tilde{T}_Q$ are defined to be range filters.
Specifically, for each node $u \in \Tilde{T}_Q$, the filter $\psi_u$ can be expressed as a list of intervals:
\begin{equation}
    \psi_u = [[l_{u,1}, h_{u,1}], \dots, [l_{u,n}, h_{u,n}]].
\end{equation}
We further say that $\psi_u$ is an \emph{empty} range filter (or an \emph{empty range} for short) if and only if there exists $1\le i\le n$ for which $l_{u,i}>h_{u,i}$.
A tuple $t \in R$ satisfies $\psi_u$ if and only if $x_i \in [l_{u,i}, h_{u,i}]$ holds for each $x_i \in \mathrm{att}(R)$.
In Example~\ref{example}, let $\psi=[[1,2],[1,4],[1,4]]$, we have $R|_\psi=\{(1,2),(2,3)\}$, $S|_\psi=S$, $T|_\psi=\{(4,2)\}$, and then $\mathit{Res}(Q_\Delta|_\psi)=\{(2,3,4)\}$.
Moreover, the filter of the root node $r$ is
\begin{equation}
\psi_r=\left[[\mathrm{min}_Q(x_1),\mathrm{max}_Q(x_1)],\dots,[\mathrm{min}_Q(x_n),\mathrm{max}_Q(x_n)]\right],
\end{equation}
where for each $1\le i\le n$, $\max_Q(x_i)=\max\mathrm{dom}_Q(x_i)$ and $\min_Q(x_i)=\min\mathrm{dom}_Q(x_i)$.
These bounds are computed during the index construction phase in $O(|Q|)$ time.
Then, once the index is constructed, $\psi_r$ can be calculated in $O(1)$ time, and $Q|_{\psi_r}=Q$ since all tuples in $\bigcup_{R\in Q}R$ satisfy $\psi_r$.

We further define an important class of range filters, say the ``\emph{prefix range filters}", on which the upper-bound algorithm and the children exploration algorithm can be calculated efficiently.
\begin{definition}
    For any range filter $\psi=[[l_1,h_1],\dots,[l_n,h_n]]$, if there exists an integer $1\le s\le n$ such that $\psi$ satisfies
    \begin{enumerate}
        \item $\forall 1\le i< s$, $l_{i}=h_{i}$,
        \item $l_{s}\le h_{s}$, and
        \item $\forall s\le i\le n$, $l_{i}=\min_Q(x_i),h_{i}=\max_Q(x_i)$,
    \end{enumerate}
    then $\psi$ is a \emph{prefix range filter} with a \emph{split position} $s$.
    Moreover, if $s+1$ is not a split position of $\psi$, then $s$ is the \emph{maximum split position} of $\psi$.
\end{definition}
It is clear that $\psi_r$ is a prefix range filter with split position $1$.
We will later discuss the advantage properties of the prefix range filters and prove that all filters in the RRATree are prefix range filters.

\subsection{The Upper-Bound Algorithm}\label{SUBSEC:upp}
Let $G_Q$ be the schema graph of $Q$, given a fractional edge cover $c\in EC(G_Q)$, for the sake of ease in theoretical analysis,
we initially define $upp(\psi)=\lfloor\textrm{AGM}_c(Q|_\psi)\rfloor$ for any range filter $\psi$.
Since all the sizes of relations are integers, the floor of the AGM bound still serves as an upper-bound for the size of the join result.
Then it holds immediately that $upp(\psi)\ge |\mathit{Res}(Q|_\psi)|$.
In Section~\ref{SEC:OPTIMIZE}, we introduce other upper bounds of the number of join results.
Tighter upper bounds will enhance the practical efficiency of the algorithm, \emph{i.e.}, reducing both the enumeration delay and the total running time.

Deng et al.~\cite{Sample2023} build a range tree $T_R$ as an index for each relation $R$ with $|\mathrm{att}(R)|=d$ in $O(|R|\log^{d-1}|R|)$ time in the index construction phase,
then for any range filter $\psi$, $|R|_\psi|$ can be computed by $T_R.count(\psi)$ in $O(\log^{d-1}|R|)$ time.
We now demonstrate that, for any prefix range filter $\psi$, the AGM bound of the filtered query $Q|_\psi$ can be computed in $O(\log |Q|)$ time. Since the AGM bound can be computed in $O(1)$ time once the cardinalities $|R|_\psi|$ for each $R \in Q$ are obtained, it suffices to show that these cardinalities can be computed within the stated time bounds.

For each $R\in Q$ with attributes $\mathrm{att}(R)=\{x_{r_1},\dots,x_{r_d}\}$
where $d=|\mathrm{att}(R)|$ and $r_1,\dots,r_d\in\mathbb{N}[1,n]$.
Assume that all tuples in $R$ are ordered lexicographically.
In particular, $\forall t,t'\in R,t\prec t'$ iff $\left(t(x_{r_1}),\dots,t(x_{r_d})\right)\prec\left(t'(x_{r_1}),\dots,t'(x_{r_d})\right)$ in lexicographical order.
Then we define the functions $R.lower:\mathbb{N}^{d}\rightarrow\mathbb{N}$ and $R.upper:\mathbb{N}^{d}\rightarrow\mathbb{N}$.
For any $d$-ary tuple $t$ (which may not necessarily be a member of $R$),
$R.lower(t)$ returns the index of the first tuple $t'\in R$ such that $t'\ge t$,
and $R.upper(t)$ returns the index of the first tuple $t'\in R$ such that $t'>t$.
Define $R[t_l,t_h]=\{t\in R|t_l\preceq t\preceq t_h\}$, and
\begin{equation}
    R.cnt(\psi)=|R[t^\psi_l,t^\psi_h]|=R.upper(t^\psi_h)-R.lower(t^\psi_l),
\end{equation}
where $\psi=[[l_1,h_1],\dots,[l_n,h_n]]$, $t^\psi_l=(l_{r_1},\dots,l_{r_d})$ and $t^\psi_h=(h_{r_1},\dots,h_{r_d})$, then the following lemma holds.

\begin{lemma}
    For any prefix range filter $\psi$, $|R|_\psi|=R.cnt(\psi)$.
\end{lemma}
\ifshowproof
\begin{proof}
    For each tuple $t\in R$ satisfying $\psi$, we have $l_{i}\le t(x_{i})\le h_i$ for $\forall i\in\{r_1,\dots,r_d\}$. This implies $t_l^\psi\preceq t\preceq t_h^\psi$, and consequently $t\in R[t_l^\psi,t_h^\psi]$. Therefore, $R|_\psi\subseteq R[t_l^\psi,t_h^\psi]$.
    For each tuple $t\in R[t_l^\psi,t_h^\psi]$, we have $t_l^\psi\preceq t\preceq t_h^\psi$.
    We need to prove that for each $x_i\in\mathrm{att}(R)$, $l_i\le t(x_i)\le h_i$.
    Let $s$ be a split position of $\psi$, then:
    \begin{enumerate}
        \item For $\forall x_i\in\mathrm{att}(R)$ with $\forall 1\le i < s$, since $l_{i}=h_{i}$, it follows that $l_i=t(x_i)=h_i$. Otherwise, one of the conditions $t\succeq t_l^\psi$ or $t\preceq t_h^\psi$ would be violated.
        \item If $x_s\in\mathrm{att}(R)$ then $l_s\le t(x_s)\le h_s$, otherwise $t_l^\psi\preceq t\preceq t_h^\psi$ would be violated.
        \item For $\forall x_i\in\mathrm{att}(R)$ with $\forall s< i \le n$, since $l_{i}=\min_Q(x_i)$ and $h_{i}=\max_Q(x_i)$, it follows that $l_i\le t(x_i)\le h_i$.
    \end{enumerate}
    Then $t$ satisfies $\psi$, \textit{i.e.}, $t\in R|_\psi$. Therefore, $R[t_l^\psi,t_h^\psi]\subseteq R|_\psi$ and then we have $R[t_l^\psi,t_h^\psi]=R|_\psi$, which implies that $R.cnt(\psi)=|R|_\psi|$.
\end{proof}
\fi

We now present the index for each relation table $R \in Q$ that supports $O(\log|R|)$-time computation of $R.cnt$.
A straightforward approach to calculate $R.cnt$ is to maintain a lexicographically sorted array of all tuples in $R$.
Given such an array, for any prefix range filter $\psi$,
both $R.lower(t_l^\psi)$ and $R.upper(t_h^\psi)$ can be calculated in $O(\log |Q|)$ time via a binary search,
then $R.cnt(\psi)$ can be calculated in $O(\log |Q|)$ time.
                            
However, this approach is inefficient under updates:
each insertion or deletion may require shifting up to $|R|$ elements to restore the sorted array, incurring $O(|R|)$ time per update.
To support both efficient query and update operations, we adopt a self-balancing binary search tree (BST), \emph{e.g.}, an AVL tree, as the underlying index structure.
Each node in the tree corresponds one-to-one to a tuple in the relation table and maintains the binary search property: the tuple in each node is lexicographically greater than that in its left child and less than that in its right child, if such children exist.
Furthermore, each node $u$ stores the size of the subtree rooted in itself, denoted as $u.size$,
which enables efficient counting of tuples within a specified range in the relation.
As both tree balancing and subtree size maintenance can be performed in $O(\log|R|)$ time per insertion or deletion, the index structure efficiently supports dynamic update with logarithmic overhead.
Leveraging the BST structure and the stored sizes, $R.lower(t)$ and $R.upper(t)$ can be calculated in $O(\log|R|)$ time for any tuple $t\in\mathbb{N}^d$ by performing a root-to-leaf traversal.
That is, for any prefix range filter $\psi$, $R.cnt(\psi)=R.upper(t_h^\psi)-R.lower(t_l^\psi)$ can be calculated in $O(\log|R|)$ time. Then we have
\begin{lemma}
    For any prefix range filter $\psi$ and any fractional edge cover $c\in EC(G_Q)$, both $\mathrm{AGM}_c(Q|_\psi)$ and $\mathrm{AGM}(Q|_\psi)$ can be calculated in $O(\log |Q|)$ time.
\end{lemma}
Therefore, for any prefix range filter $\psi$, $upp(\psi)=\lfloor\textrm{AGM}_c(Q|_\psi)\rfloor$
can be computed in $O(\log|Q|)$ time.


We now aim to prove that when $upp(\psi)\le 1$, the join result $Res(Q|_\psi)$ can be computed in $O(\log|Q|)$ time. To this end, we first define a property of $upp$, called super-additivity, and prove that when $upp$ satisfies the property, $Res(Q|_\psi)$ can be computed in $O(\log|Q|)$ time for any $\psi$ satisfying $upp(\psi)\le 1$.

\begin{property}[super-additivity]\label{prop:upp-superadditivity}
  Given any range filter $\psi = [[l_1, h_1], \dots, [l_n, h_n]]$ and $1 \leq p \leq n$,
  for any partition of the interval $[l_p, h_p]$ into $k$ disjoint sub-intervals $I_1, \dots, I_k$ such that $[l_p, h_p] = \bigcup_{i=1}^k I_i$ and $I_i \cap I_j = \emptyset$ for $i \neq j$, the following inequality holds:
    \begin{equation}
        \sum_{i=1}^k upp([[l_1,h_1],\dots,I_i,\dots,[l_n,h_n]])\le upp(\psi).
    \end{equation}
\end{property}

If an upper-bound algorithm satisfies Property~\ref{prop:upp-superadditivity}, it is referred to as \emph{super-additive}. Then the following lemma holds.

\begin{lemma}\label{LEM:AGMLEONE}
    If $upp$ is super-additive, then for any prefix range filter $\psi = [[l_{1}, h_{1}], \dots, [l_{n}, h_{n}]]$ such that $upp(\psi)\le 1$, $\mathit{Res}(Q|_{\psi})$ can be computed in $O(\log|Q|)$ time.
\end{lemma}

\ifshowproof
\begin{proof}
Since $upp(\psi)\le 1$, two cases need to be discussed.
Firstly,
for the case $upp(\psi)=0$,
we have $|\mathit{Res}(Q|_{\psi})|\le upp(\psi)=0$,
which implies that $\mathit{Res}(Q|_{\psi})=\emptyset$.
Secondly, 
for the case $upp(\psi)=1$,
we have $|\mathit{Res}(Q|_{\psi})|\leq1$.
Let $s=\max\{i|l_i\neq h_i\}$,
by Property~\ref{prop:upp-superadditivity},
there exists at most one integer $p\in\mathbb{N}[l_{s}, h_{s}]$ such that
\begin{itemize}
    \item $upp([[l_{1},h_{1}],\dots,[l_{s},p-1],\dots,[l_{n},h_{n}]])=0$,
    \item $upp([[l_{1},h_{1}],\dots,[p,p],\dots,[l_{n},h_{n}]])=1$,
    \item $upp([[l_{1},h_{1}],\dots,[p+1,h_s],\dots,[l_{n},h_{n}]])=0$.
\end{itemize}
If $upp([[l_{1},h_{1}],\dots,[p,p],\dots,[l_{n},h_{n}]])=0$ follows for every $p\in\mathbb{N}[l_{s},h_{s}]$, we have $\mathit{Res}(Q|_{\psi})=\emptyset$.
Otherwise, it is able to calculate the only integer $p\in\mathbb{N}[l_{s},h_{s}]$ by a multi-head binary search (introduced in Section~\ref{SEC:CHILDREN}) in $O(\log|Q|)$ time.
Then apply the same procedure to the next unfixed position of $[[l_{1},h_{1}],\dots,[p,p],\dots,[l_{n},h_{n}]]$, and continue this process iteratively on other positions as needed. If this process finally obtains $\psi^*=[[l_{1}^*, h_{1}^*], \dots, [l_{n}^*, h_{n}^*]]$ such that $upp(\psi^*)=1$ and $\forall 1\le i\le n,l_i^*=h_i^*$, then $(l_1^*,\dots,l_n^*)$ is the only tuple in $\mathit{Res}(Q|_\psi)$.
Otherwise, we have $\mathit{Res}(Q|_{\psi})=\emptyset$.
The time it takes to decide whether $\mathit{Res}(Q|_\psi)=\emptyset$ and calculate $\psi^*$ (if exists) is at most $O(n\log|Q|)$, which is $O(\log|Q|)$ in data complexity.
\end{proof}
\fi





Moreover, the upper-bound algorithm proposed in this section is super-additive.
This can be readily derived from the AGM Split Theorem~\cite{Sample2023}, which states that the AGM bound is super-additive over interval partitions.
Hence, for any prefix range filter $\psi$ with $upp(\psi) \le 1$, $\mathit{Res}(Q|_\psi)$ can be computed in $O(\log |Q|)$ time.


\subsection{The Children Exploration Algorithm} \label{SEC:CHILDREN}
During the path-finding process of \textsc{RRAccess}, in order to calculate the children of each visited node in $O(\log|Q|)$ time, we develop an efficient children exploration algorithm, which takes a prefix range filter $\psi$ as input and outputs at most a constant number of children filters $\psi_1,\dots,\psi_k$ as required in Definition~\ref{DEF:RRATREE}.
Our approach follows the active domain partitioning strategy proposed by Deng et al.~\cite{Sample2023}.
We improve their method by introducing a novel partitioning algorithm and more efficient data structures, which together reduce the computational complexity from $O(\log^d|Q|)$ to $O(\log|Q|)$, where $d=\max_{R\in Q}\mathrm{att}(R)$ denotes the maximum arity of the relations.

\textbf{The divide operation.}
Specifically, for any range filter $\psi=[[l_1,h_1],\dots,[l_n,h_n]]$, let $s=\min\{i|l_i\neq h_i\}$, an operation named ``divide'' is defined, which takes the filter $\psi$ as input and outputs:
\begin{itemize}
    \item $\psi_{\mathrm{left}}=[[l_1,h_1],\dots,[l_s,p-1],\dots,[l_n,h_n]]$,
    \item $\psi_{\mathrm{mid}}=[[l_1,h_1],\dots,[p,p],\dots,[l_n,h_n]]$,
    \item $\psi_{\mathrm{right}}=[[l_1,h_1],\dots,[p+1,h_s],\dots,[l_n,h_n]]$
\end{itemize}
such that
\begin{enumerate}
    \item $upp(\psi_\mathrm{left})+upp(\psi_\mathrm{mid})+upp(\psi_\mathrm{right})\le upp(\psi)$,
    \item $upp(\psi_\mathrm{left})\le\frac{1}{2}upp(\psi),upp(\psi_\mathrm{right})\le\frac{1}{2}upp(\psi)$.
\end{enumerate}

Assuming the upper-bound algorithm is super-additive,
then (1) holds for any $p\in\mathbb{N}[l_s,h_s]$.
Deng et al.~\cite{Sample2023} calculate the proper division point $p$ that satisfies (2) by a binary search for the minimized integer in $[l_s,h_s]$ such that $upp(\psi_\mathrm{left})\ge \frac{1}{2}upp(\psi)$.
In each iteration of their binary search process, a count oracle implemented by a range tree is employed to calculate the cardinality of each filtered relation. This results in a time complexity of $O(\log^d|Q|)$ for the divide operation, where $d=\max_{R\in Q}|\mathrm{att}(R)|$.
In contrast, we show that when dividing a prefix range filter, the division point $p$ can be computed in $O(\log |Q|)$ time.
We achieve this by first establishing that $s$ is a valid split position of $\psi$, then reducing the problem to an equivalent optimization problem, and finally devising an efficient $O(\log |Q|)$-time algorithm to solve it.

\begin{lemma}
    Let $\psi=[[l_1,h_1],\dots,[l_n,h_n]]$ be a prefix range filter, then $s=\min\{i|l_i\neq h_i\}$ is a split position of $\psi$.
\end{lemma}
\begin{proof}
    Otherwise, there exists $s<s'\le n$ such that $[l_{s'},h_{s'}]\neq [\min_Q(x_{s'}),\max_Q(x_{s'})]$, then for $\forall i \neq s'$, $i$ is not a split position of $\psi$. Moreover, $s'$ is not a split position since $l_s\neq h_s$, then $\psi$ is not a prefix range filter, which leads to a conflict.
\end{proof}
For each relation $R_i\in Q$ satisfying $x_s\in\mathrm{att}(R_i)$, define $A_i =\left[t_1(x_s),\dots,t_{n_i}(x_s)\right]$,
where $n_i=|R_i|_\psi|$, $R|_\psi=\{t_1,\dots,t_{n_i}\}$ and $t_1\preceq\dots\preceq t_{n_i}$.
We observe that the task of ``calculating the division point $p$" in the divide operation can be cast as an instance of the following optimization problem:
\begin{description}
    \item[\textbf{Input:}] a constant number of sorted integer arrays $A_1,\dots,A_k$ and an integer $T$,
    \item[\textbf{Output:}] the maximum integer $p^*$ such that $F(p^*)\le T$, 
\end{description}
where $N_i(p)=|\{x\in A_i|x<p\}|$, and $F:\mathbb{N}^k\rightarrow\mathbb{N}$ is an $k$-ary non-decreasing function (For simplicity, we use $F(p)$ as shorthand for $F(N_1(p),\dots,N_k(p))$).
The reduction is obvious since
$upp$ can be viewed as a non-decreasing function over the cardinalities of the filtered relations $R_{\mathrm{left}}$ for each relation $R\in Q$ with $x_s \in \mathrm{att}(R)$.


Assuming for each relation table $R\in Q$, all tuples in $R$ are sorted lexicographically,
we propose the Multi‑Head Binary Search (MHBS) algorithm (see Algorithm~\ref{ALG:MHBS}), which solves the above problem in $O(\log \sum_{i=1}^k|A_i|)$ time.
Intuitively, the MHBS algorithm maintains a binary search interval for each array, and in each iteration, it halves the interval of one selected array, until the intervals converge, yielding the optimal result.

\begin{algorithm}
    \caption{Multi Head Binary Search (MHBS)}\label{ALG:MHBS}
    \LinesNumbered
    \KwIn{$A_1,\dots,A_k$, $T$}
    \KwOut{maximum $p^*$ such that $F(p^*)\le T$}
    \lIf{$F(|A_1|,\dots,|A_k|)\le T$}{\Return{$+\infty$}}
    \For{$1\le i\le k$}{
        \If{$r_i-l_i\le 1$}{
            \lIf{$F(A_{i}[l_i]+1)>T$}{\Return{$A_{i}[l_{i}]$}}
            \lElse{$m_{i}\leftarrow r_i$}
        }
        \lElse{
        $l_i\leftarrow 0$, $r_i\leftarrow|A_i|$, $m_i\leftarrow\lfloor\frac{l_i+r_i}{2}\rfloor$}
    }
    \While{$\exists i,r_i-l_i>1$}{
        $i_{\min}\leftarrow\arg\min\limits_{r_i-l_i>1}A_i[m_i]$\;
        $i_{\max}\leftarrow\arg\max\limits_{r_i-l_i>1}A_i[m_i]$\;
        \lIf{$F(m_1,\dots,m_k)\le T$}{
            $l_{i_{\min}}\leftarrow m_{i_{\min}}$
        }
        \lElse{
            $r_{i_{\max}}\leftarrow m_{i_{\max}}$
        }
        \lFor{$i\in\{i_{\min},i_{\max}\}$}{$m_{i}\leftarrow\lfloor\frac{l_{i}+r_{i}}{2}\rfloor$}
        \If{$\exists i\in\{i_{\min},i_{\max}\},r_{i}-l_{i}\le 1$}{
            \lIf{$F(A_{i}[l_i]+1)>T$}{\Return{$A_{i}[l_i]$}}
            \lElse{$m_{i}\leftarrow r_{i}$}
        }
    }
    \Return{$\min\limits_{1\le i\le k}A_i[r_i]$};
\end{algorithm}

\begin{lemma}
    Algorithm~\ref{ALG:MHBS} returns the maximum integer $p^*$ such that $F(p^*)\le T$ in $O(\log \sum_{i=1}^k|A_i|)$ time.
\end{lemma}

\ifshowproof

\begin{proof}
For the convenience of the proof, a dummy element $+\infty$ is appended to the end of each sequence, that is, $A_i[|A_i|]=+\infty$ for each $1\le i\le k$.
If $F(|A_1|,\dots,|A_k|)\le T$, the algorithm obviously returns $p^*$.
Then for the case where $F(|A_1|,\dots,|A_k|)> T$, we show that Algorithm~\ref{ALG:MHBS} maintains the invariant
\begin{equation}\label{EQ:INV}
A_i[l_i]\le p^*\le A_i[r_i],\quad \forall\,1\le i\le k,
\end{equation}
and then prove that Algorithm~\ref{ALG:MHBS} correctly returns $p^*$ after at most $O(\log\sum_{i=1}^k|A_i|)$ iterations.

\emph{Base case.}  Initially, for each $1\le i\le k$ we set $l_i=0$ and $r_i=|A_i|$. Since each $A_i$ is sorted,
\begin{equation}
A_i[0]\le\min A_i\le p^*\le\max A_i\le A_i[n],
\end{equation}
then the invariant holds before the first iteration.

\emph{Inductive step.}  Suppose at the start of some iteration we have $A_i[l_i]\le p^*\le A_i[r_i]$ for any $1\le i\le k$.
We aim to demonstrate that the invariant is preserved after the update performed in the current iteration.
Without loss of generality, we assume that for any $1\le i\le k$, $r_i - l_i > 1$ satisfies. (If some $i$ already has $r_i - l_i \le 1$ and the algorithm has not yet terminated, then by the invariant we have $A_i[l_i] < p^* \le A_i[r_i]$ and $N_i(A_i[r_i]) \le N_i(A_i[l_i]+1) = r_i$,
then $N_i(p^*) = r_i = m_i$, and the problem reduces to the remaining $k-1$ sequences.)
In each iteration, the algorithm modifies exactly one of the two bounds: it either updates the lower bound $l_{i_{\min}}$ for some index $i_{\min}$, or updates the upper bound $r_{i_{\max}}$ for some index $i_{\max}$.  We discuss the two cases:

\textbf{Case 1:} $F(m_1,\dots,m_k)\le T$.

Let $p=A_{i_{\min}}[m_{i_{\min}}]=\min_{r_i - l_i > 1} A_i[m_i]$.
Since
\begin{equation}
N_{i_{\min}}(p)
=\bigl|\{x\in A_{i_{\min}}\mid x < A_{i_{\min}}[m_{i_{\min}}]\}\bigr|
\le m_{i_{\min}},
\end{equation}
and for each $i\neq i_{\min}$,
\begin{equation}
N_i(p)
\le N_i\bigl(A_i[m_i]\bigr)
=\bigl|\{x\in A_i\mid x < A_i[m_i]\}\bigr|
\le m_i,
\end{equation}
it follows that
\begin{equation}
F\bigl(N_1(p),\dots,N_k(p)\bigr)
\le F(m_1,\dots,m_k)\le T.
\end{equation}
Hence $p^*\ge p$, then after updating $l_{i_{\min}}\leftarrow m_{i_{\min}}$, we still have $A_{i_{\min}}[l_{i_{\min}}]\le p^*\le A_{i_{\min}}[r_{i_{\min}}]$.
For all $i\neq i_{\min}$, neither $l_i$ nor $r_i$ changes, so the invariant is preserved after the update.

\textbf{Case 2:} $F(m_1,\dots,m_k)> T$.

Let $p=A_{i_{\max}}[m_{i_{\max}}]=\max_{r_i - l_i > 1} A_i[m_i]$.
Since
\begin{equation}
    N_{i_{\max}}(p+1)=|\{x\in A_{i_{\max}}|x\le A_{i_{\max}}[m_{i_{\max}}]\}|\ge m_{i_{\max}},
\end{equation}
and for each $i\neq i_{\min}$,
\begin{equation}
    N_i(p+1)\ge N_i(A_i[m_i]+1)=|\{x\in A_i|x\le A_i[m_i]\}|\ge m_i,
\end{equation}
it follows that
\begin{equation}
    F(N_1(p+1),\dots,N_k(p+1))\ge F(m_1,\dots,m_k)> T.
\end{equation}
Thus $p^*<p+1$ which implies $p^*\le p$, then after updating $r_{i_{\max}} \leftarrow m_{i_{\max}}$, we still have $A_{i_{\max}}[l_{i_{\max}}]\le p^*\le A_{i_{\max}}[r_{i_{\max}}]$.
For all $i\neq i_{\max}$, neither $l_i$ nor $r_i$ changes, so the invariant is preserved after the update.

\emph{Convergence and termination}.
In each iteration, an update is performed on sequence $A_i$ only if $r_i - l_i > 1$,
and every such update reduces the length of the interval by at least one.
Consequently, the length of each interval $[l_i,r_i]$ shrinks to at most $2$ after at most $O(\log |A_i|)$ updates on $A_i$.
Since the number of sequences is a constant, the total number of iterations of the algorithm is bounded by $O(\log \sum_{i=1}^k|A_i|)$.

Now we show that when Algorithm~\ref{ALG:MHBS} terminates, it returns $p^*$ correctly. If it
 terminates at Line 4 or Line 14, which implies
\begin{equation}
F\bigl(N_1(A_{i}[l_i]+1),\dots,N_k(A_{i}[l_{i}]+1)\bigr) > T,
\end{equation}
it follows that $p^*\le A_{i}[l_i]$, then by the invariant $p^*\ge A_i[l_i]$, we have $p^* = A_i[l_i]$.

If Algorithm~\ref{ALG:MHBS} terminates at line 16, which implies $r_i - l_i \le 1$ and
$F(N_1(A_i[l_i]+1),\dots,N_k(A_i[l_i]+1))\le T$ satisfies for every $1\le i\le k$.
Let $p=\min_{1\le i\le k}A_i[r_i]$ and $q=\max_{1\le i\le k}A_i[l_i]$, then
\begin{equation}
    \begin{aligned}
F(N_1(p),\dots&,N_k(p))\le F(N_1(A_1[r_1]),\dots,N_k(A_k[r_k]))\\
&\le F(N_1(A_1[l_1]+1),\dots,N_k(A_k[l_k]+1))\\
&\le F(N_1(q+1),\dots,N_k(q+1))\le T.
\end{aligned}
\end{equation}
Thus $p^*\ge p$, and by the invariant, $p^*\le p$, we conclude $p^* =p= \min_{1\le i\le k}A_i[r_i]$.

In summary, since $i_{\min}$, $i_{\max}$, and $F$ can be computed in $O(1)$ time, and computing each $N_i$ takes at most $O(\log |A_i|)$ time, Algorithm~\ref{ALG:MHBS} returns $p^*$ in $O(\log\sum_{i=1}^k|A_i|)$ time.
\end{proof}
\fi

Since $\sum_{i=1}^k|A_i|\le\sum_{i=1}^k|R_i|\le |Q|$, the following corollary follows immediately.

\begin{corollary}
    The divide operation on any prefix range filter can be performed in $O(\log |Q|)$ time.
\end{corollary}

\emph{Remarks.} The assumption that all tuples in each relation table are sorted lexicographically is made for analytical convenience and does not limit generality, as other hierarchical indexing structures (such as balanced trees, B-trees, skip lists, and tries) can be supported with minor modifications in a manner similar to that discussed in Section~\ref{SUBSEC:upp}.

Then for any prefix range filter $\psi$, $children(\psi)$ can be calculated in a recursive way in $O(\log|Q|)$ time, as shown in Algorithm~\ref{ALG:CHILDREN}.

\begin{algorithm}
    \caption{$children$}\label{ALG:CHILDREN}
    \LinesNumbered
    \KwIn{$\psi$}
    \KwOut{a list of filters}
    \lIf{$upp(\psi)\le 1$}{\textbf{return} $\emptyset$}
    $res\leftarrow\emptyset$\;
    divide $\psi$ into $\psi_\mathrm{left},\psi_\mathrm{mid},\psi_\mathrm{right}$\;
    \If{$upp(\psi_\mathrm{left})>0$ and $\psi_{\textrm{left}}$ is not empty}{$res\leftarrow res\cup\{\psi_\mathrm{left}\}$}
    \lIf{$upp(\psi_\mathrm{mid})=1$}{$res\leftarrow res\cup\{\psi_\mathrm{mid}\}$}
    \lElse{$res\leftarrow res\cup children(\psi_\mathrm{mid})$}
    \If{$upp(\psi_\mathrm{right})>0$ and $\psi_{\textrm{right}}$ is not empty}{$res\leftarrow res\cup\{\psi_\mathrm{right}\}$}
    \textbf{return} res\;
\end{algorithm}

\begin{figure}[t]
\centering
\begin{subfigure}[h]{0.12\textwidth}
\centering
\begin{tikzpicture}
  \coordinate (A) at (0,0,0);
  \coordinate (B) at (1.5,0,0);
  \coordinate (C) at (1.5,1.5,0);
  \coordinate (D) at (0,1.5,0);
  \coordinate (E) at (0,0,1.5);
  \coordinate (F) at (1.5,0,1.5);
  \coordinate (G) at (1.5,1.5,1.5);
  \coordinate (H) at (0,1.5,1.5);
  \coordinate (x) at (0.75,0,1.5);
  \coordinate (y) at (1.5,0,0.75);
  \coordinate (z) at (1.5,0.75,0);
  \coordinate (psi) at (0.75,1.5,0);

  \node at (x) [below] {x};
  \node at (y) [below right] {y};
  \node at (z) [right] {z};
  \node at (psi) [above] {$\psi$};

  \draw (B) -- (C) -- (D);
  \draw [dashed] (A) -- (B);
  \draw [dashed] (A) -- (D);
  \draw (E) -- (F) -- (G) -- (H) -- cycle;
  \draw [dashed] (A) -- (E);
  \draw (B) -- (F);
  \draw (C) -- (G);
  \draw (D) -- (H);

\end{tikzpicture}
\end{subfigure}
\begin{subfigure}[h]{0.1\textwidth}
\centering
    \begin{tikzpicture}
        \draw[->, thick] (0,0) -- (1.3,0);
        \node at (0.65,0) [above] {children};
        \node at (0.65,0) [below] { };
    \end{tikzpicture}
\end{subfigure}
\begin{subfigure}[h]{0.21\textwidth}
\centering
\begin{tikzpicture}
  \coordinate (A) at (0-\eps,0,0);
  \coordinate (B) at (1.5+1.5*\eps,0,0);
  \coordinate (C) at (1.5+1.5*\eps,1.5,0);
  \coordinate (D) at (0-\eps,1.5,0);
  \coordinate (E) at (0-\eps,0,1.5);
  \coordinate (F) at (1.5+1.5*\eps,0,1.5);
  \coordinate (G) at (1.5+1.5*\eps,1.5,1.5);
  \coordinate (H) at (0-\eps,1.5,1.5);

  \coordinate (LB) at (0.75-\eps,0,0);
  \coordinate (LC) at (0.75-\eps,1.5,0);
  \coordinate (LF) at (0.75-\eps,0,1.5);
  \coordinate (LG) at (0.75-\eps,1.5,1.5);

  \coordinate (RB) at (0.75+1.5*\eps,0,0);
  \coordinate (RC) at (0.75+1.5*\eps,1.5,0);
  \coordinate (RF) at (0.75+1.5*\eps,0,1.5);
  \coordinate (RG) at (0.75+1.5*\eps,1.5,1.5);

  \coordinate (MB) at (0.75,0,0-0.5*\eps);
  \coordinate (MC) at (0.75,1.5,0-0.5*\eps);
  \coordinate (MF) at (0.75,0,1.5+0.5*\eps);
  \coordinate (MG) at (0.75,1.5,1.5+0.5*\eps);

  \coordinate (MBB) at (0.75,0,0.75-0.5*\eps);
  \coordinate (MCC) at (0.75,1.5,0.75-0.5*\eps);
  \coordinate (MFF) at (0.75,0,0.75+0.5*\eps);
  \coordinate (MGG) at (0.75,1.5,0.75+0.5*\eps);

  \coordinate (MMU) at (0.75,0-0.2*\eps,0.75);
  \coordinate (MMUU) at (0.75,0.75-0.2*\eps,0.75);
  \coordinate (MMD) at (0.75,1.5+0.2*\eps,0.75);
  \coordinate (MMDD) at (0.75,0.75+0.2*\eps,0.75);

  \coordinate (MMM) at (0.75,0.75,0.75);

  \coordinate (x) at (1.125+1.5*\eps,0,1.5);
  \coordinate (y) at (1.5+1.5*\eps,0,0.75);
  \coordinate (z) at (1.5+1.5*\eps,0.75,0);
  \coordinate (psi1) at (0.375-\eps,1.5,0);
  \coordinate (psi2) at (0.75,0,1.5);
  \coordinate (psi3) at (0.75,1.5,0.75);
  \coordinate (psi4) at (0.75,0.75,0.75);
  \coordinate (psi5) at (0.75,0,0.75);
  \coordinate (psi6) at (0.75,1.5,0);
  \coordinate (psi7) at (1.125+1.5*\eps,1.5,0);


    \fill [green!30, fill opacity=0.5] (D) -- (LC) -- (LG) -- (H) --cycle;
    \fill [green!40, fill opacity=0.5] (E) -- (LF) -- (LG) -- (H) --cycle;
    \fill [green!50, fill opacity=0.5] (LB) -- (LF) -- (LG) -- (LC) --cycle;

    \fill [green!30, fill opacity=0.5] (RC) -- (RG) -- (G) -- (C) --cycle;
    \fill [green!40, fill opacity=0.5] (RG) -- (G) -- (F) -- (RF) --cycle;
    \fill [green!50, fill opacity=0.5] (C) -- (B) -- (F) -- (G) --cycle;
    
  \draw (LB) -- (LC) -- (D);
  \draw [dashed] (A) -- (LB);
  \draw [dashed] (A) -- (D);
  \draw (E) -- (LF) -- (LG) -- (H) -- cycle;
  \draw (LB) -- (LF);
  \draw (LC) -- (LG);

  \draw (B) -- (C) -- (RC);
  \draw [dashed] (RB) -- (B);
  \draw [dashed] (RB) -- (RC);
  \draw (RF) -- (F) -- (G) -- (RG) -- cycle;
  \draw [dashed] (RB) -- (RF);
  \draw (RC) -- (RG);
  
  \draw [dashed] (A) -- (E);
  \draw (B) -- (F);
  \draw (C) -- (G);
  \draw (D) -- (H);

  \fill [blue!30, fill opacity=0.5] (MB) -- (MC) -- (MCC) -- (MBB) -- cycle;
  \fill [blue!30, fill opacity=0.5]  (MF) -- (MG) -- (MGG) -- (MFF) -- cycle;
  \draw (MB) -- (MC) -- (MCC) -- (MBB) -- cycle;
  \draw (MF) -- (MG) -- (MGG) -- (MFF) -- cycle;

  \draw [orange] (MMU) -- (MMUU);
  \draw [orange] (MMD) -- (MMDD);

  \fill (MMM) circle (1pt);
  
  \node at (x) [below] {x};
  \node at (y) [below right] {y};
  \node at (z) [right] {z};
  \node at (psi1) [above] {$\psi_1$};
  \node at (psi2) [below] {$\psi_2$};
  \node at (psi3) [above] {$\psi_3$};
  \node at (psi4) [right] {$\psi_4$};
  \node at (psi5) [below] {$\psi_5$};
  \node at (psi6) [above] {$\psi_6$};
  \node at (psi7) [above] {$\psi_7$};

\end{tikzpicture}

\end{subfigure}
\caption{Devision of $\psi$ of $Q_\Delta$.}
\label{FIG:children}
\end{figure}
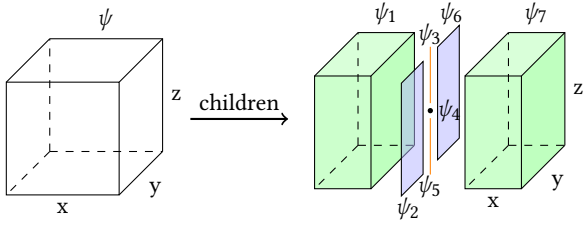

Since the recursive depth is at most $d$, Algorithm~\ref{ALG:CHILDREN} returns up to $2d+1\le O(1)$ filters.
In Example~\ref{example}, a prefix range filter $\psi$ may be split into at most $7$ filters, as illustrated in Figure~\ref{FIG:children}.
Let $upp(\psi)=\lfloor\mathrm{AGM}_{c^*}(Q|_\psi)\rfloor$ for any filter $\psi$ where $c^*(R)=c^*(S)=c^*(T)=\frac{1}{2}$, $\Tilde{T}_{Q_\Delta}$ is finally constructed as shown in Figure~\ref{fig:agmc}.
Moreover, by the definition of the division operation and Property~\ref{prop:upp-superadditivity}, it is easy to prove that the returned filters satisfy the properties in Definition~\ref{DEF:RRATREE}.

To establish that Algorithm~\ref{ALG:RRAccess} is a relaxed random-access algorithm running in $O(\log^2|Q|)$ worst-case time and $O(\log|Q|)$ amortized time, it remains to show that all range filters in $\Tilde{T}_Q$ are prefix range filters.

\begin{lemma}
    For any prefix range filter $\psi$, let $\psi_{\mathrm{left}}$, $\psi_{\mathrm{mid}}$ and $\psi_{\mathrm{right}}$ denote the three filters obtained by dividing $\psi$, then all non‑empty ranges among these filters are prefix range filters.
\end{lemma}
\ifshowproof
\begin{proof}
    Let $\psi=[[l_1,h_1],\dots,[l_n,h_n]]$, since $s=\min\{i|l_i\neq h_i\}$ is a split position of $\psi$, it follows that
    \begin{enumerate}
        \item $\forall 1\le i<s$, $l_i=h_i$,
        \item $\forall s<i\le n$, $l_i=\min_Q(x_i)$ and $h_i=\max_Q(x_i)$.
    \end{enumerate}
    If $\psi_{\mathrm{left}}$ is not an empty range, then $l_p\le mid_p-1$, which implies that $\psi_{\mathrm{left}}$ is a prefix range filter with split position $s$.
    Similarly, if $\psi_{\mathrm{left}}$ is not an empty range, then $mid_p+1\le h_p$, which implies that $\psi_{\mathrm{right}}$ is a prefix range filter with split position $s$.
    Finally, since $mid_p\le mid_p$, $\psi_{\mathrm{mid}}$ is a prefix range filter with split position $s$.
\end{proof}
\fi
Since the filter of the root node, $\psi_r$, is a prefix range filter, we can conclude by induction that

\begin{corollary}
    For any node $u\in\Tilde{T}_Q$, $\psi_u$ is a prefix range filter.
\end{corollary}

\textbf{The Near-Optimal \textsc{REnum} Algorithm.} The aforementioned implementations give a theoretical guarantee of the performance of \textsc{REnum}, thus yielding the following theorem.
\begin{theorem}
\label{mainthm}
There exists a constructive random-order enumeration algorithm for join queries with expected $O(\frac{\mathrm{AGM}(Q)}{|\mathit{Res}(Q)|+1}\log^2|Q|)$ delay and $O(\mathrm{AGM}(Q)\log|Q|)$ total running time, after an $O(|Q|\log|Q|)$-time index construction phase.
\end{theorem}
\begin{proof}

\begin{figure}[t]
    \centering
    \includegraphics[width=\linewidth, page=1]{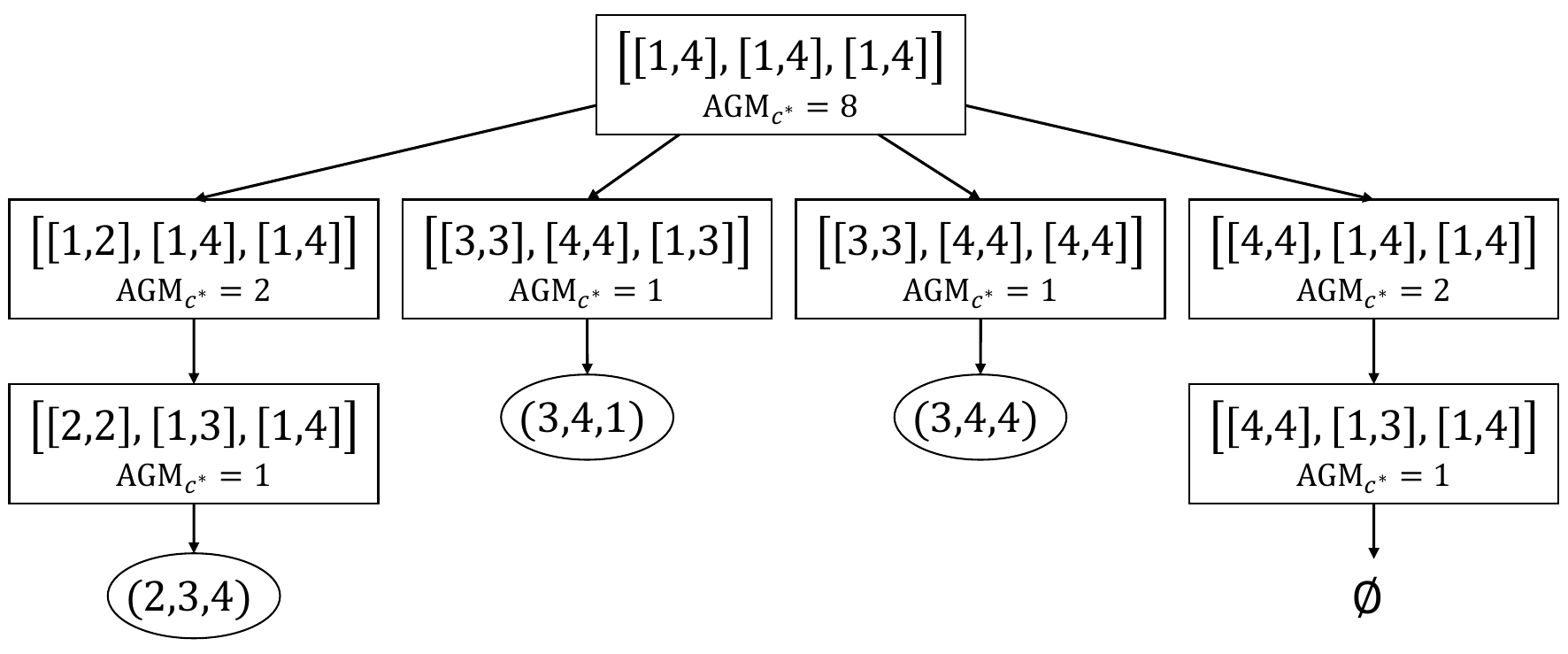}
    \caption{$\Tilde{T}_{Q_\Delta}$ when $upp(\psi)=\lfloor\mathrm{AGM}_{c^*}(Q_\Delta|_\psi)\rfloor$}
    \label{fig:agmc}
\end{figure}
We prove the theorem by analyzing Algorithm~\ref{ALG:BPREnum} under the setting where $N=upp(\psi_r)$ and $upp(\psi)=\left\lfloor\mathrm{AGM}_{c^*}(Q|_\psi)\right\rfloor$ for any range filter $\psi$.
First, the indexes of the relation tables can be built in $O(|Q|\log |Q|)$ time.
After the indexes have been built, the computation of $c^*$ and $\psi_r$ takes a constant time.
Then, according to Lemma~\ref{lem:upperbound}, $\varphi^*_Q(i) = \textit{false}$ for each $i>upp(\psi_r)$.
Since $upp(\psi)=\left\lfloor\mathrm{AGM}_{c^*}(Q|_\psi)\right\rfloor\le\mathrm{AGM}(Q)$,
by Lemma~\ref{LEM:RRAccess}, Algorithm~\ref{ALG:RRAccess} is a relaxed random-access algorithm running in $O(\log^2|Q|)$ worst-case time and $O(\log|Q|)$ amortized time.
Lastly, since $N=upp(\psi_r)$,
by Lemma~\ref{lem:RREnum}, the expected enumeration delay of Algorithm~\ref{ALG:BPREnum} is $O(\frac{upp(\psi_r)}{|\mathit{Res}(Q)|+1}\log^2|Q|)=O(\frac{\mathrm{AGM}(Q)}{|\mathit{Res}(Q)|+1}\log^2|Q|)$, and its total running time is at most $O(\mathrm{AGM}(Q)\log|Q|)$.
\end{proof}
Theorem~\ref{thmm} and the results from~\cite{Sample2023} demonstrate that, under the combinatorial $k$-clique hypothesis, both the expected delay and the total running time of Algorithm~\ref{ALG:BPREnum} are nearly worst-case optimal, after an $O(|Q| \log |Q|)$-time index construction.

\section{Speed Up the Enumeration}\label{SEC:OPTIMIZE}
It is easy to see that the larger the ratio $\frac{\mathrm{AGM}(Q)}{|Res(Q)|+1}$,
the longer the enumeration delay and the total running time of \textsc{REnum}.
To address this bottleneck, based on our enumeration framework, we propose two speed-up techniques that significantly improve its efficiency in practice, especially for queries where $\mathrm{AGM}(Q)\gg |Res(Q)|$.


\subsection{Larger Trivial Interval Discovery}\label{SEC:LTI}
We observe that trivial intervals often appear as consecutive sequences and can thus be grouped into larger intervals, particularly when $\mathrm{AGM}(Q)\gg|Res(Q)|$.
To avoid the frequent picking of such trivial integers during enumeration, we propose the Larger Trivial Interval discovery (LTI) technique.
In contrast to the previous implementation of \textsc{RRAccess}, which returns only a single-point interval $[i, i]$ when $\varphi_Q(i) = \textit{false}$, LTI discovers larger trivial intervals, so that more trivial integers are prevented from being picked in the follow-up steps, thereby speeding up the enumeration.

\textbf{The trivial intervals of filters.}
Observe that when $\varphi_Q^*(i)=false$, Algorithm~\ref{ALG:RRAccess} may return a trivial interval only at line 5 or line 11.
By the definition of $\Tilde{T}_Q$, for each filter $\psi$ satisfying $children(\psi)=\psi_1,\dots,\psi_k$, we have $upp(\psi)\ge \sum_{i=1}^kupp(\psi_i)$.
Then if an integer $i$ (after removing the offset) falls within $(\sum_{i=1}^kupp(\psi_i),upp(\psi)]$, it follows that $\varphi^*_Q(i)=\textit{false}$.
That is, such intervals should be discovered and banned in the follow-up steps.
Then for line 5, since for any integer $\mathit{offset}+\sum_{j=1}^kupp(\psi_k)< t\le \mathit{offset}+upp(\psi)$ we have $\varphi^*_Q(t)=\textit{false}$, we define the trivial interval of the filter $\psi$ as
\begin{equation}
    I_\psi=\left[\mathit{offset}+\sum_{j=1}^kupp(\psi_k)+1,\mathit{offset}+upp(\psi)\right],
\end{equation}
and let \textsc{RRAccess}$^{Q,\varphi^*}$ return $I_\psi$ instead of $[i,i]$.
For line 11, since the algorithm has already reached a leaf node of $\Tilde{T}_Q$ (\emph{i.e.} $upp(\psi)=1$), and the leaf node corresponds to no result tuple, we define the trivial interval of $\psi$ as $I_\psi=[i, i]$ and let \textsc{RRAccess}$^{Q,\varphi^*}$ return it.

In Example~\ref{example}, when we call \textsc{RRAccess}$^{Q_\Delta,\varphi^*}(7)$ in the enumeration process, the algorithm first computes $C_r\leftarrow children(\psi_r)$.
As shown in Figure~\ref{fig:agmc}, $\sum_{\psi\in C_r}upp(\psi)=6<7$, thus it returns $I_{\psi_r}=\left[\sum_{\psi\in C_r}upp(\psi)+1,upp(\psi_r)\right]=[7,8]$.
Then the Ban-Pick tree bans $[7,8]$ and stops picking the integers in it during the follow-up steps, thereby eliminating the need to call \textsc{RRAccess}$^{Q_\Delta,\varphi^*}(8)$.

Notice that each trivial interval $[l, h]$ can be returned by \textsc{RRAccess}$^{Q,\varphi^*}$ at most once during the enumeration process.
Otherwise, some integer $i \in [l, h]$ would be selected after $[l, h]$ has been banned, leading to a contradiction.
Moreover, by the definition of $\Tilde{T}_Q$ and $\varphi^*$, the trivial intervals of the filters in $\Tilde{T}_Q$ are disjoint from each other.
We evaluate this basic LTI technique in Section~\ref{SEC:EXPERIMENTS}.

\textbf{Merging contiguous trivial intervals.}
We find that many trivial intervals are contiguous and can be merged to form a larger interval.
Specifically, by the definition of $\Tilde{T}_Q$ and $\varphi^*$, for any filter $\psi$ in $\Tilde{T}_Q$, let $\psi'\in children(\psi)$ be the last child of $\psi$,
then the intervals $I_\psi$ and $I_{\psi'}$ are contiguous.
For example, in $\Tilde{T}_{Q_\Delta}$ (Figure~\ref{fig:agmc}), the last child of the root filter $\psi_r$ is $\psi_r'=[[4,4],[1,4],[1,4]]$,
then the trivial interval $I_{\psi_r'}=[6,6]$ is contiguous with $I_{\psi_r}=[7,8]$.
Moreover, once we obtain a trivial interval $I_\psi$ of a filter $\psi$ in the recursive process of \textsc{RRAccess}, to compute such contiguous trivial intervals that can be merged with $I_\psi$, one can recursively visit the last child of the current filter, and determine whether they can be merged with their parents' trivial intervals upon the recursive return (for instance, if $\psi$ is the last child of its parent $\psi^*$, then the trivial interval $I_{\psi^*}$ can be merged with $I_\psi$).
Then we get the merged trivial interval in $O(\log|Q|)$ time, without calling \textsc{RRAccess} too many times.
As a result, the number of calls to \textsc{RRAccess} is reduced during the enumeration, leading to better practical performance.
Moreover, it can be proven that during the enumeration process, no two merged trivial intervals overlap.
We evaluate this variant of basic LTI (say merging trivial intervals, MTI) in Section~\ref{SEC:EXPERIMENTS}.

\textbf{Batch Trivial Interval Discovery.}
To further reduce the number of calls to \textsc{RRAccess}, we enhance the MTI technique to discover and report multiple trivial intervals within a single run of \textsc{RRAccess}.
Specifically, along the top-to-down traversal path of the RRATree, we compute the trivial intervals of each visited node and recursively explore the chain of its last child.
All the discovered trivial intervals along this path are then merged (when possible) and reported to the Ban-Pick Tree for banning.
In this way, although the enumeration delay increases to $O(\frac{\mathrm{AGM(Q)}}{|Res(Q)|}\log^3|Q|)$, the total running time remains $O(\mathrm{AGM}(Q)\log|Q|)$, and the number of calls to \textsc{RRAccess} is reduced during the enumeration.
We evaluate this variant of MTI (say batch trivial interval discovery, BTI) in Section~\ref{SEC:EXPERIMENTS}.

\subsection{Tighter Upper-Bound Estimation}\label{SEC:TU}
This subsection introduces the Tighter Upper-bound estimation (TU) technique, which enhances the effectiveness of LTI and makes more and larger trivial intervals discovered and banned earlier.
We begin by illustrating how TU improves the efficiency of enumeration through a motivating example.

\begin{figure}[t]
    \centering
    \includegraphics[width=\linewidth, page=2]{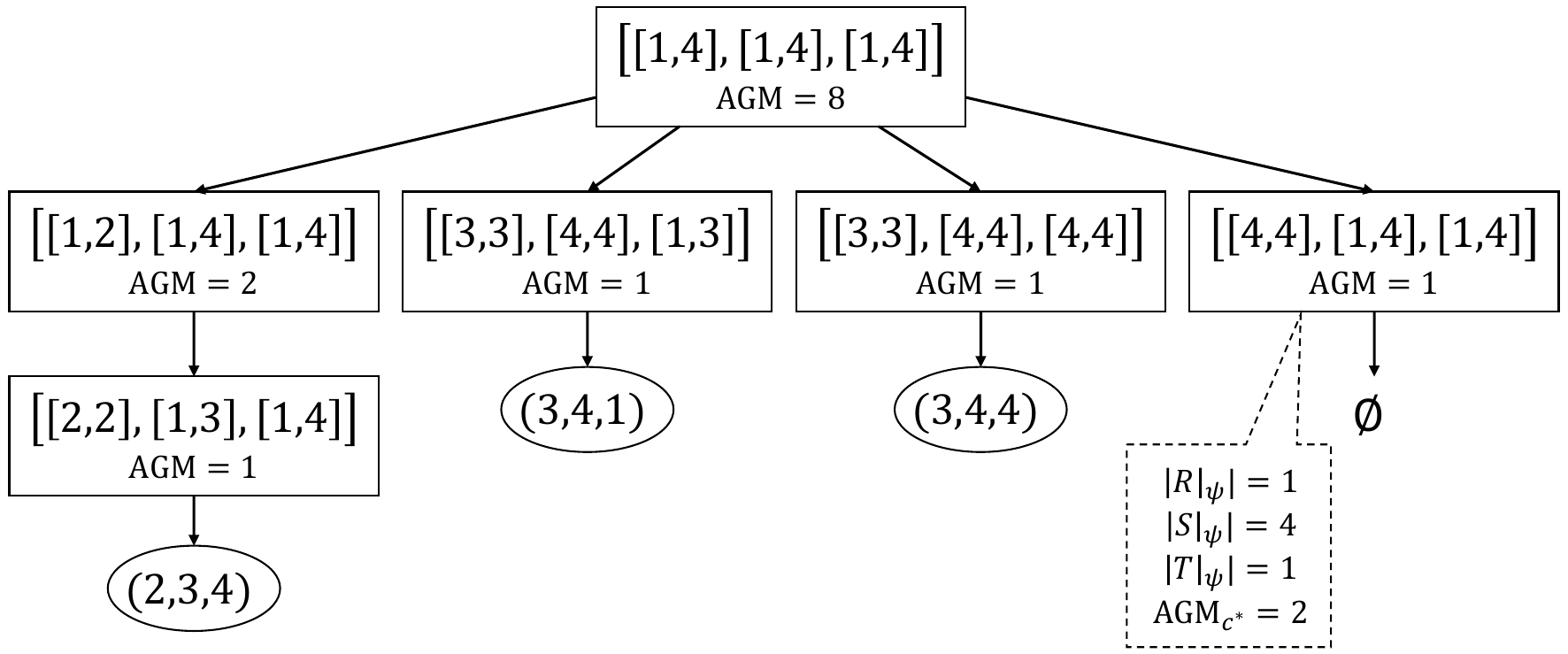}
    \caption{$\Tilde{T}_{Q_\Delta}$ when $upp(\psi)=\lfloor\mathrm{AGM}(Q_\Delta|_\psi)\rfloor$}
    \label{fig:agm}
\end{figure}

Let $upp(\psi)=\lfloor\mathrm{AGM}(Q|_\psi)\rfloor$ instead of $\lfloor\mathrm{AGM}_{c^*}(Q|_\psi)\rfloor$ for any $\psi$, then $\Tilde{T}_{Q_\Delta}$ is constructed as Figure~\ref{fig:agm}.
For $\psi=[[4,4],[1,4],[1,4]]$, we have $|R|_\psi|=|T|_\psi|=1$ and $|S|_\psi|=4$.
Therefore $upp(\psi)=\lfloor\mathrm{AGM}(Q_\Delta|_\psi)\rfloor=1<\lfloor\mathrm{AGM}_{c^*}(Q_\Delta|_\psi)\rfloor=2$.
This tighter upper bound increases the size of the trivial interval of $\psi_r$, yielding $I_{\psi_r} = [6,8]$ instead of $[7,8]$.
As a result, TU expands the trivial intervals of filters in the lower levels of the RRATree,
which makes more trivial integers to be discovered and banned earlier,
thereby enhancing the effectiveness of LTI.

In this subsection, we introduce two different upper-bound algorithms. It can be verified that both of them satisfy the properties in Definition~\ref{DEF:RRATREE} and are super-additive.
In practice, for each prefix range filter $\psi$, we compute the minimum value among these upper-bound algorithms as the estimated upper bound of $|Res(Q|_\psi)|$, say $upp^*(\psi)$.
In this way, it is clear that $upp^*$ satisfies the properties defined in Definition~\ref{DEF:RRATREE}.
Moreover, it is super-additive, as formalized in the following lemma.
\begin{lemma}
    If the upper-bound algorithms $upp_1,\dots,upp_c$ are super-additive, then the upper-bound algorithm $upp^*$ where $upp^*(\psi)=\min_{i=1}^cupp_i(\psi)$ for any range filter $\psi$ is also super-additive.
\end{lemma}

\begin{proof}
    Let $\psi=[[l_1,h_1],\dots,[l_n,h_n]]$ be any prefix range filter with split position $s$.
    Let $I_1,\dots,I_k$ be a partition of the interval $[l_s,h_s]$.
    For $\forall1\le i\le k$, let $\psi_i=[[l_1,h_1],\dots,I_i,\dots,[l_n,h_n]]$, then
    \begin{equation}
        \sum_{i=1}^kupp(\psi_i)\le\sum_{i=1}^kupp_{i^*}(\psi_i)\le upp^*_{i^*}(\psi)=upp^*(\psi),
    \end{equation}
    where $i^*=\arg\min_{i=1}^kupp_i(\psi)$. 
\end{proof}

That is, $upp^*$ can be used as a replacement for the upper-bound algorithm presented in Section~\ref{SEC:RRATree}.

\textbf{Upper-bound based on minimized AGM bound.} 
Given any join query $Q\in\mathcal{Q}$, $|Res(Q)|\le\lfloor\mathrm{AGM}(Q)\rfloor$.
During the enumeration process, the minimal AGM bound $\mathrm{AGM}(Q|_\psi)$ can be calculated by solving a linear program in $O(1)$ time after computing the sizes of relations $\{|R|_\psi||R\in Q\}$ in $O(\log|Q|)$ time.

However, solving a linear program to compute the minimized AGM bound incurs significant computational overhead, which may degrade the overall performance.
As a compromise, we heuristically select a small number of representative fractional edge covers and calculate the minimum of their corresponding AGM bounds.
This approach effectively presents a tighter upper bound while avoiding the cost of repeatedly solving linear programs.

In Example~\ref{example},
during the top-down traversal of $\Tilde{T}_{Q_\Delta}$, the active domain of each attribute is progressively reduced in order $x$, $y$, and $z$.  
Since $x$ is the first attribute to be constrained,
the cardinalities of the filtered relations involving $x$, \emph{i.e.}, $R$ and $T$, will decrease more rapidly in each root-to-leaf path of $\Tilde{T}_{Q_\Delta}$. 
Therefore, higher fractional edge cover weights of $R$ and $T$ lead to a smaller AGM bound.
For instance, consider the prefix range filter $\psi = [[4,4], [1,4], [1,4]]$.
Let $c\in EC(Q_\Delta)$ satisfy $c(R)=c(T)=1$ and $c(S)=0$, then we have $\mathrm{AGM}_c(Q_\Delta|_\psi)=1<\mathrm{AGM}_{c^*}(Q_\Delta|_\psi)=2$.
This example demonstrates that assigning larger fractional edge cover weights to relations involving attributes that appear earlier in the domain-reduction order can lead to a tighter AGM bound at lower levels of the RRATree.
In practice, we heuristically select a small constant number of fractional edge covers based on this principle, and ensure that for every relation $R \in Q$, there exists at least one selected cover $c \in EC(Q)$ such that $c(R) >0$.


\textbf{Upper-bound based on acyclic skeleton query.} Any cyclic join query can be transformed into an acyclic one by removing a subset of relations~\cite{Sample2018}.
Specifically, given any cyclic join query $Q$, it can be decomposed into two subqueries, $Q_s$ and $Q_r$, such that $Q=Q_s\Join Q_r$.
Here, $Q_s$ is an acyclic query (referred to as the \emph{skeleton query}) and $Q_r$ is the query consisting of the remaining relations (referred to as the \emph{residual query}).
In Example~\ref{example}, the triangle query $Q_\Delta$ can be decomposed into an acyclic skeleton subquery $Q_{\Delta s}=R(x,y)\Join S(y,z)$ and a residual subquery $Q_{\Delta r}=T(x,z)$.
Based on the above decomposition, following lemma holds:

\begin{lemma}
    For any join query $Q$ and filter $\psi$, if $\mathrm{att}(Q_s) = \mathrm{att}(Q)$, then $|Res(Q|_\psi)|\le |Res(Q_s|_\psi)|$. Otherwise, if $\mathrm{att}(Q_s)\subsetneq\mathrm{att}(Q)$, for any fractional edge cover $c\in EC(G_{Q_r\setminus Q_s})$, $|Res(Q|_\psi)|\le |Res(Q_s|_\psi)|\cdot\mathrm{AGM}_c(Q_r^*|_\psi)$,
    where $Q_r^*=\{R[\mathrm{att}(Q_r)\setminus\mathrm{att}(Q_s)]|R\in Q_r\}$, and for any attribute set $V\subseteq\mathrm{att}(R)$, $R[V]=\{t[V]|t\in R\}$.
\end{lemma}
\ifshowproof
\begin{proof}
    If $\mathrm{att}(Q_s) = \mathrm{att}(Q)$, then $Res(Q|_\psi)\subseteq Res(Q_s|_\psi)$, which implies $|Res(Q|_\psi)|\le |Res(Q_s|_\psi)|$.
    If $\mathrm{att}(Q_s)\subsetneq\mathrm{att}(Q)$, then
    \begin{equation}
        \begin{aligned}
            |Res(Q|_\psi)|&=|Res(Q_s|_\psi)\Join Res(Q_r|_\psi)|\\
            &\le \sum_{t\in Res(Q_s|_\psi)}|Res((Q_r\ltimes t)|_\psi)|\\
            &\le |Res(Q_s|_\psi)|\max_{t\in Res(Q_s|_\psi)}|Res((Q_r\ltimes t)|_\psi)|\\
            &\le |Res(Q_s|_\psi)|\cdot|Res(Q_r^*|_\psi)|\\
            &\le |Res(Q_s|_\psi)|\cdot\mathrm{AGM}_c(Q_r^*|_\psi).
        \end{aligned}
    \end{equation}
\end{proof}
\fi
We build the indexes described in Section~\ref{SEC:RRATree} on $Q_r^*$ before the enumeration, so that $\mathrm{AGM}_c(Q_r^*|_\psi)$ can be computed in $O(\log|Q|)$ time for any $\psi$.
For $|Res(Q_s|_\psi)|$,
we assume that each relation $R = \{t_1, \dots, t_{|R|}\}$ is ordered lexicographically, and calculate an array $A_R$ and its prefix-sum array for each $R\in Q_s$, such that:
\begin{equation}
    \forall 1\le i\le |R|,\quad A_R[i]=\left|\mathop{\Join}\limits_{R' \in T_R}R'\ltimes t_i\right|,
\end{equation}
where $T_R$ denotes the subtree of the join tree of $Q_s$ rooted at $R$.
These arrays can be computed using a dynamic programming approach similar to that described in~\cite{Sample2018},
which takes $O(|Q|\log |Q|)$ time before the enumeration.
Then, for each table $R$ and any prefix range filter $\psi$,
the cardinality $|(\Join_{R' \in T_R} R') \Join R|_\psi|$ can be efficiently computed in $O(\log |R|)$ time with the help of a prefix-sum array of $A_R$.
Subsequently, $|Res(Q_s|\psi)|$ can be computed in $O(1)$ time using these cardinalities for all $R \in Q$.
Although this method requires an $O(|Q|\log|Q|)$-time query-specific preprocessing, it can effectively reduce the enumeration delay for many queries in practice, especially in the early stages of the enumeration.

We implement these upper-bound algorithms, referred to as A (minimized AGM-based) and S (Skeleton-based), and evaluate their effectiveness in speeding up the enumeration in Section~\ref{SEC:EXPERIMENTS}.

\section{Experiments}\label{SEC:EXPERIMENTS}

In this section, we present an experimental evaluation of our random-order enumeration algorithm for join queries along with the speed-up techniques.
We denote the basic random-order enumeration algorithm introduced in Section~\ref{SEC:OVERVIEW} as \textsc{REnum}.
We evaluate the optimized variants of \textsc{REnum}, each combining an LTI technique from Section~\ref{SEC:LTI} with a TU technique from Section~\ref{SEC:TU}.
These variants are denoted uniformly as \textsc{REnum}\textsubscript{X-Y}, where $\text{X} \in \{\text{L},\text{M}, \text{B}\}$ indicates the use of the basic LIT, MTI or BTI, and $Y \in \{\text{A}, \text{AS}\}$ indicates the employed upper-bound algorithm.


\begin{figure*}[t!]
    \centering
    \begin{subfigure}[b]{0.24\textwidth} \label{totalq2}
        \centering
        \includegraphics[width=\textwidth]{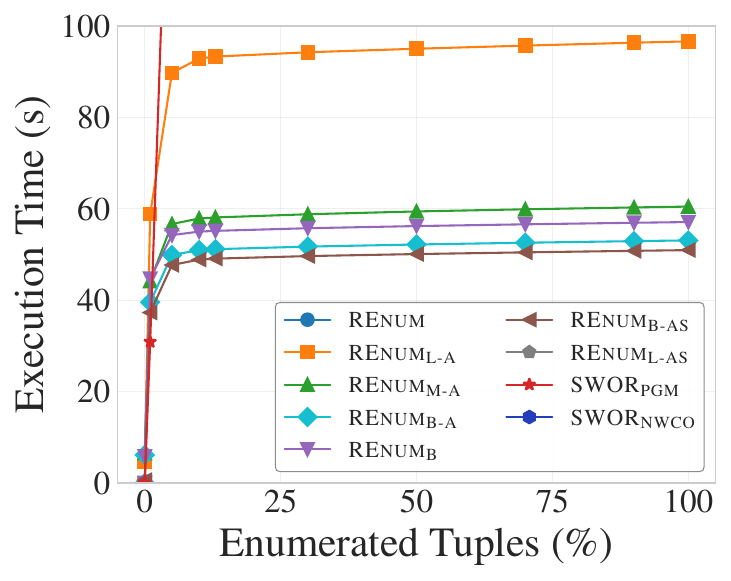}
        \caption{$Q_A$}
    \end{subfigure}
    \hfill
    \begin{subfigure}[b]{0.24\textwidth} \label{totalq3}
        \centering
        \includegraphics[width=\textwidth]{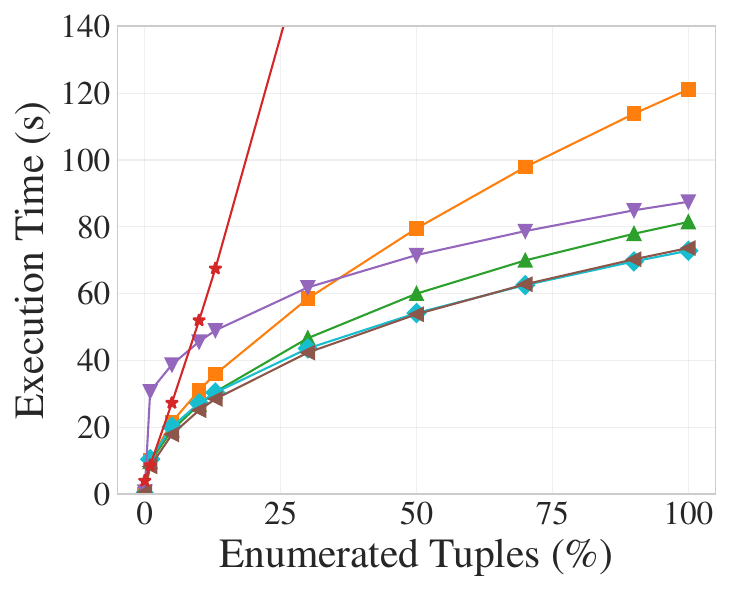}
        \caption{$Q_\Delta$}
    \end{subfigure}
    \hfill
    \begin{subfigure}[b]{0.24\textwidth} \label{totalq4}
        \centering
        \includegraphics[width=\textwidth]{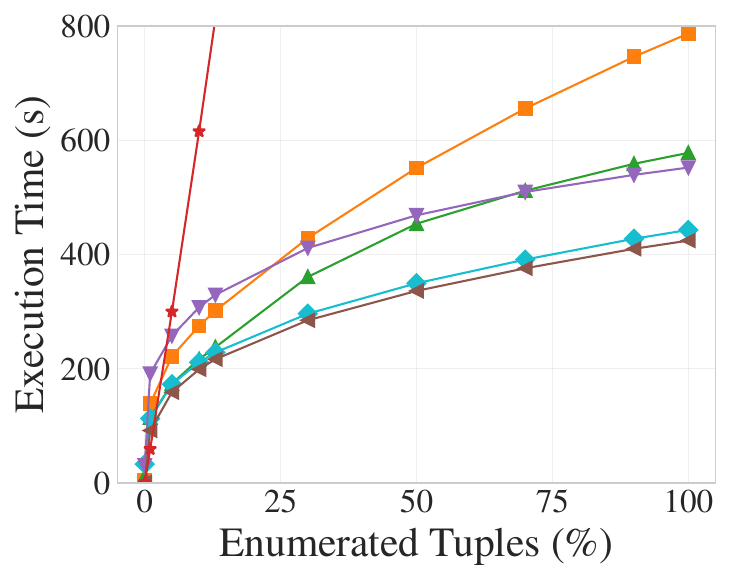}
        \caption{$Q_S$}
    \end{subfigure}
    \hfill
    \begin{subfigure}[b]{0.24\textwidth} 
        \centering
        \includegraphics[width=\textwidth]{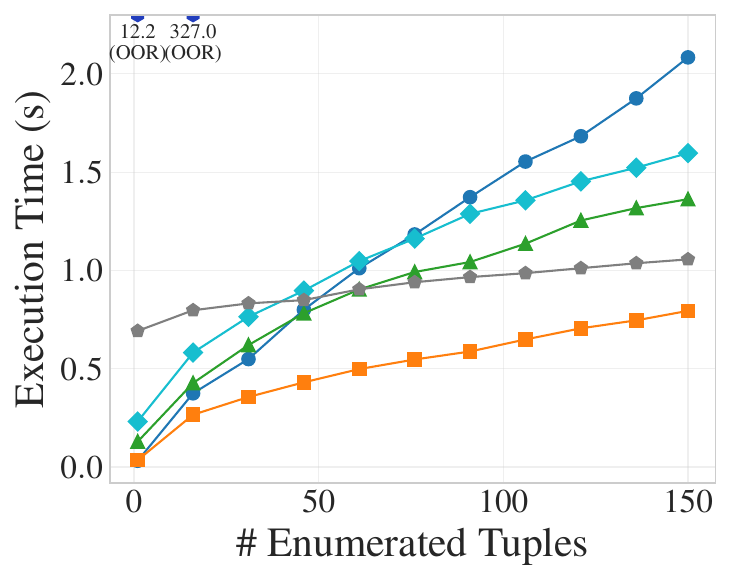}
        \caption{$Q_\Delta$ (early stages)}
    \end{subfigure}
    \caption{Total enumeration time}
    \label{fig:EXPTIME}
\end{figure*}

\begin{figure*}[t!]
    \centering
    \begin{subfigure}[b]{0.28\textwidth} 
        \centering
        \includegraphics[height=12em]{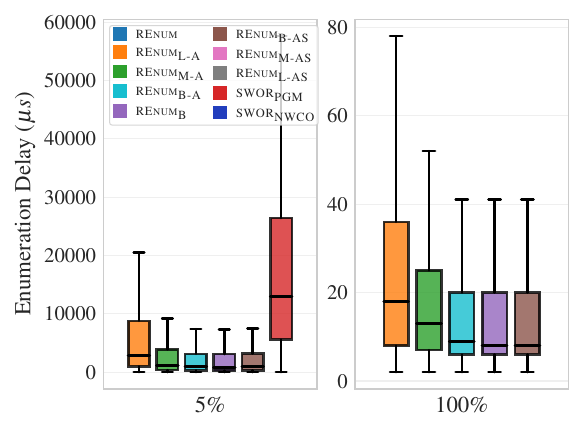}
        \caption{$Q_A$}
    \end{subfigure}
    \hfill
    \begin{subfigure}[b]{0.28\textwidth} 
        \centering
        \includegraphics[height=12em]{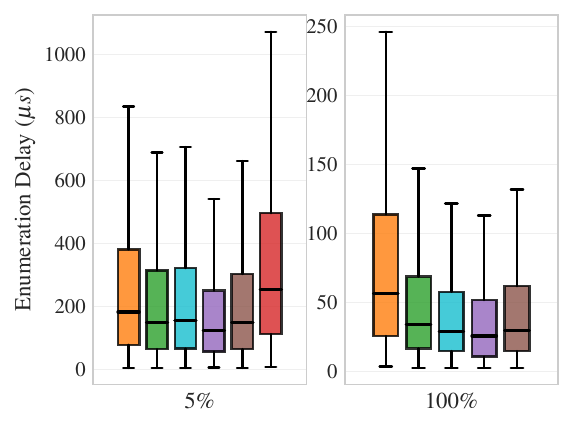}
        \caption{$Q_\Delta$}
    \end{subfigure}
    \hfill
    \begin{subfigure}[b]{0.28\textwidth} 
        \centering
        \includegraphics[height=12em]{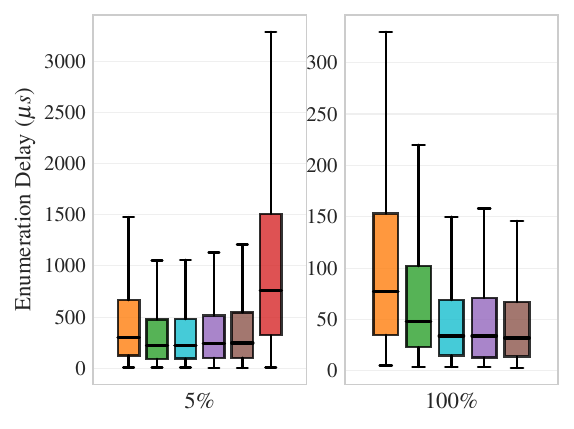}
        \caption{$Q_S$}
    \end{subfigure}
    \hfill
    \begin{subfigure}[b]{0.14\textwidth} 
        \centering
        \includegraphics[height=12em]{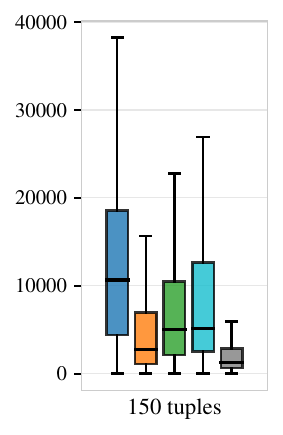}
        \caption{$Q_\Delta$ (early stages)}
    \end{subfigure}
    \caption{Enumeration delay}
    \label{fig:EXPDELAY}
\end{figure*}
\subsection{Experimental Setup}
We now describe the setup of our experiments.

\textbf{Algorithms.} 
We compare our algorithms with sampling-based methods derived from two join sampling algorithms:

\begin{enumerate}
    \item \textbf{PGMJoin}~\cite{pgmjoin}: the state-of-the-art among all existing implemented algorithms (with preprocessing), which requires $\Omega(|Q|)$-time preprocessing for each query.
    \item \textbf{NWCO}~\cite{Sample2023}: a theoretically nearly worst-case optimal join sampling algorithm without query-specific preprocessing.
\end{enumerate}
We use the implementation of PGMJoin from their public repository. As there is no publicly available implementation of NWCO, we implemented it for our experiments.
These algorithms generate uniform samples with replacement. We adapt them to a sampling without replacement (SWOR) process by naively discarding duplicate results that have already been seen.
We denote the resulting variants as \textsc{SWOR}\textsubscript{PGM} and \textsc{SWOR}\textsubscript{NWCO}, respectively.

\textbf{Datasets.}
In our experiments, we use two datasets: \textbf{TPC-DS+} and \textbf{Twitter}.
TPC-DS+ is an extension of the standard TPC-DS benchmark (with a scale factor of $5$), augmented with an additional customer-to-customer relation table with $3\times10^5$ tuples, that models follow relationships in a social network.
The source node of each tuple is sampled uniformly, while its target node is drawn from a normal distribution. The resulting follow graph is treated as undirected by adding reciprocal edges.
Twitter is a real-world dataset that refers to follower links and profiles of twitter users, which is also used in~\cite{twitter2010,pgmjoin,Sample2018}, and we treat the follow graph as undirected.
To avoid excessive running times during the complete performance evaluation, we uniformly sample $5$ million nodes and run the algorithms on their induced subgraphs.
All relation tables in these datasets are instantiated in memory, with the necessary indexes for the algorithms constructed in advance.

\textbf{Queries.}
We evaluate the algorithms and the speed-up techniques across three queries.
On the TPC-DS+ dataset, we evaluate $Q_A$, which returns tuples consisting of a pair of customers and $2$ items they have both purchased, under the condition that the customers are connected in the social network. 
On the Twitter dataset, we evaluate queries that return all triangles ($Q_\Delta$) and 4-cliques ($Q_S$).

\textbf{Platform and Hardware.}
All experiments are performed on a workstation running 64-bit Ubuntu22.04 LTS, equipped with one Intel Xeon E7-8860@2.2GHz, 384GB DDR4 RAM, and 2TB SSD.
    \subsection{Experimental Results}

\textbf{Total enumeration time.} To characterize the total enumeration time of our algorithms, we compare it against that of SWOR\textsubscript{PGM} and SWOR\textsubscript{NWCO} on these queries.
We record the elapsed time from the beginning to the moment after the $k$-th join result is enumerated, with $k$ represents $1\%$ to $100\%$ of the total result size.
The experimental results are presented in Figure~\ref{fig:EXPTIME} (a)-(c) with a chart per query.
As both \textsc{SWOR}\textsubscript{NWCO} and the basic \textsc{REnum} require prohibitively long time to produce even $1\%$ of the join results, we only record their performance on $Q_\Delta$ in the early stages of the enumeration, which is presented in Figure~\ref{fig:EXPTIME}~(d).

The results indicate that, with the speed-up techniques, our algorithm significantly outperforms \textsc{SWOR}\textsubscript{PGM} and \textsc{SWOR}\textsubscript{NWCO}.
As an ablation study, Figure~\ref{fig:EXPTIME}(a)–(c) show that \textsc{REnum}\textsubscript{B-A} enumerates faster than \textsc{REnum}\textsubscript{M-A}, which in turn performs better than \textsc{REnum}\textsubscript{L-A}, and all of them significantly outperform the basic algorithm \textsc{REnum}. This demonstrates the effectiveness of the basic LTI, MTI and BTI.
Similarly, the results of \textsc{REnum}\textsubscript{B}, \textsc{REnum}\textsubscript{B-A} and \textsc{REnum}\textsubscript{B-AS} verify the effectiveness of TU.
At the early stages of the enumeration, we record the total enumeration time of the first $150$ enumerated tuples, as shown in Figure~\ref{fig:EXPTIME} (d).
Note that all these tuples (except those of \textsc{SWOR}\textsubscript{NWCO}) are enumerated before \textsc{SWOR}\textsubscript{PGM} completing its preprocessing and producing its first result tuple.
The results show that our algorithm enumerates faster in the early stages without MTI or BTI, due to the overhead incurred by discovering and merging numerous trivial intervals.
Moreover, \textsc{REnum}\textsubscript{L-AS} requires less preprocessing time compared to \textsc{SWOR}\textsubscript{PGM}, and after preprocessing, its total enumeration time increases slower than that of other algorithms.
\textbf{Enumeration delay.}
Due to its poor performance, \textsc{SWOR}\textsubscript{NWCO} is excluded from the delay analysis, and the delay of \textsc{REnum} is evaluated only during the early stages (the first $150$ tuples) of enumeration on $Q_\Delta$.
We record the enumeration delay of each result tuple and analyze the delay distribution by calculating the delays for the first $5\%$ and all results (only the first $5\%$ for \textsc{SWOR}\textsubscript{PGM}), which are presented in a box plot, as shown in Figure~\ref{fig:EXPDELAY} (a)-(c).
The delays in the early stages of $Q_\Delta$ are shown in Figure~\ref{fig:EXPDELAY} (d).
Outliers that fell outside the whiskers are not shown, since some are several orders of magnitude larger than the median.



The results show that the enumeration delay of our algorithms (with the speed-up tricks) is much smaller than that of \textsc{SWOR}\textsubscript{PGM} and \textsc{SWOR}\textsubscript{NWCO}. The experiments further reveal that while MTI and BTI reduce the overall enumeration delay, they tend to increase the delay in the early stages due to the overhead of trivial interval discovery and merging. In contrast, TU mainly reduces the enumeration delay in the early stages. These findings are consistent with the total enumeration time results presented earlier.

\section{Conclusions}
This paper demonstrates that the sampling-based random-order enumeration is not the end of the story,
by presenting a more efficient random-order enumeration algorithm with a worst-case guarantee on the total running time for join queries, along with two techniques to speed up the enumeration.
Our algorithm is efficient and flexible, achieving nearly optimal expected delay and total running time in the worst case under the combinatorial $k$-clique hypothesis, without large hidden constants or query-specific preprocessing.
In our experiments, with the speed-up techniques, our algorithm significantly outperforms the state-of-the-art sampling-based methods.
This allows the join results to be produced more quickly in random order within data analysis pipelines.
An interesting future work direction is to implement and further optimize these algorithms in external-memory and distributed database systems to support queries in large-scale data analysis.

\bibliographystyle{ACM-Reference-Format}
\bibliography{main}

\end{document}